\documentclass[a4paper]{jpconf}

\usepackage{amsfonts,amsmath,amssymb}
\usepackage{graphicx,wasysym}
\usepackage{tabularx,hhline}
\usepackage[dvipsnames]{xcolor}
\usepackage{color,bbold}
\usepackage{framed}
\usepackage[square,sort&compress,numbers]{natbib}

\usepackage{hyperref} 
\hypersetup{
colorlinks = True,
linkcolor = green!50!blue,
urlcolor = violet,
citecolor = red
}

\begin{document}

\title{Flat bands and Dirac cones in breathing lattices}
\author{Karim Essafi, L D C Jaubert}
\address{Okinawa Inst Sci \& Technol, Onna, Okinawa 904 0495, Japan}

\author{M Udagawa}
\address{Department of Physics, Gakushuin University, Mejiro, Toshima-ku, Tokyo 171-8588, Japan}

\ead{karim.essafi@oist.jp}
\begin{abstract}
In breathing pyrochlores and kagomes, couplings between neighbouring tetrahedra and triangles are free to differ. Breathing lattices thus offer the possibility to explore a different facet of the rich physics of these systems. Here we consider nearest-neighbour classical Heisenberg interactions, both ferromagnetic and antiferromagnetic, and study how the anisotropy of breathing lattices modifies the mode spectrum of pyrochlore and kagome systems. The nature and degeneracy of the flat bands are shown to be preserved for any value of the anisotropy. These flat bands can coexist with Dirac nodes at the $\Gamma$ point when the model becomes particle-hole symmetric. We also derive the nature of the ground state for the breathing kagome lattice, which bears a spontaneous chirality when neighbouring triangles are alternatively ferromagnetic and antiferromagnetic.
\end{abstract}

\section{Introduction}

Pyrochlore and kagome antiferromagnets are archetypical examples of flat-band systems in frustrated magnetism~\cite{Reimers1991,Garanin99a}. In these classical spin liquids, the flat bands correspond to highly degenerate ground-state manifolds 
which support emergent Coulomb physics~\cite{Henley2010}. However, while the pyrochlore and kagome antiferromagnets have been rather extensively studied, little is known about their ``breathing'' counterparts. Pyrochlore and kagome lattices are made of corner-sharing units, respectively tetrahedra and triangles. The term ``breathing'' has been coined to describe lattices where spin couplings of neighbouring units are different~\cite{Okamoto2013a}. Such geometry has first been used as a mathematical conveniency to study the quantum spin liquid nature of the pristine spin$-1/2$ pyrochlore antiferromagnet~\cite{Harris1991,Canals1998,Tsunetsugu2001,Tsunetsugu2001a,Berg2003,Kotov2004,Kim2008}. But recently, breathing lattices have gained noticeable interest on their own as possible outcomes of spin-lattice coupling~\cite{Bzdusek2015,Aoyama2016}, and after being synthesized in the rare-earth based pyrochlore material Ba$_{3}$Yb$_{2}$Zn$_{5}$O$_{11}$~\cite{Kimura2014,Savary2015,Haku2016,Rau2016,Haku2016a}, in the vanadium oxyfluoride kagome [NH$_{4}$]$_{2}$[C$_{7}$H$_{14}$N][V$_{7}$O$_{6}$F$_{18}$] ~\cite{Aidoudi2011,Clark2013,Schaffer2016} and in the spinel oxides LiGaCr$_{4}$O$_{8}$ and LiInCr$_{4}$O$_{8}$~\cite{Okamoto2013a,Tanaka2014a,Okamoto2015,Nilsen2015,Lee2016}, where the large magnetic moment of the Cr$^{3+}$ ions makes a classical approach sensible~\cite{Benton2015,Aoyama2016}.

In this paper, our goal is to complement the understanding of classical breathing lattices by studying the evolution of the flat bands originally present in the uniform lattices. The anisotropy of breathing lattices will be used as a tunable parameter, exploring both antiferromagnetic and ferromagnetic couplings between classical Heisenberg spins. We will explain why the flat bands, and their degeneracy, are protected, and how Dirac cones appear at the $\Gamma$ point when neighbouring units have couplings of exactly opposite sign. In addition to their band structures, the ground state of the breathing kagome lattice will be described, both for the Heisenberg and XXZ Hamiltonians.

\begin{figure}
\centering
\includegraphics[width=5cm]{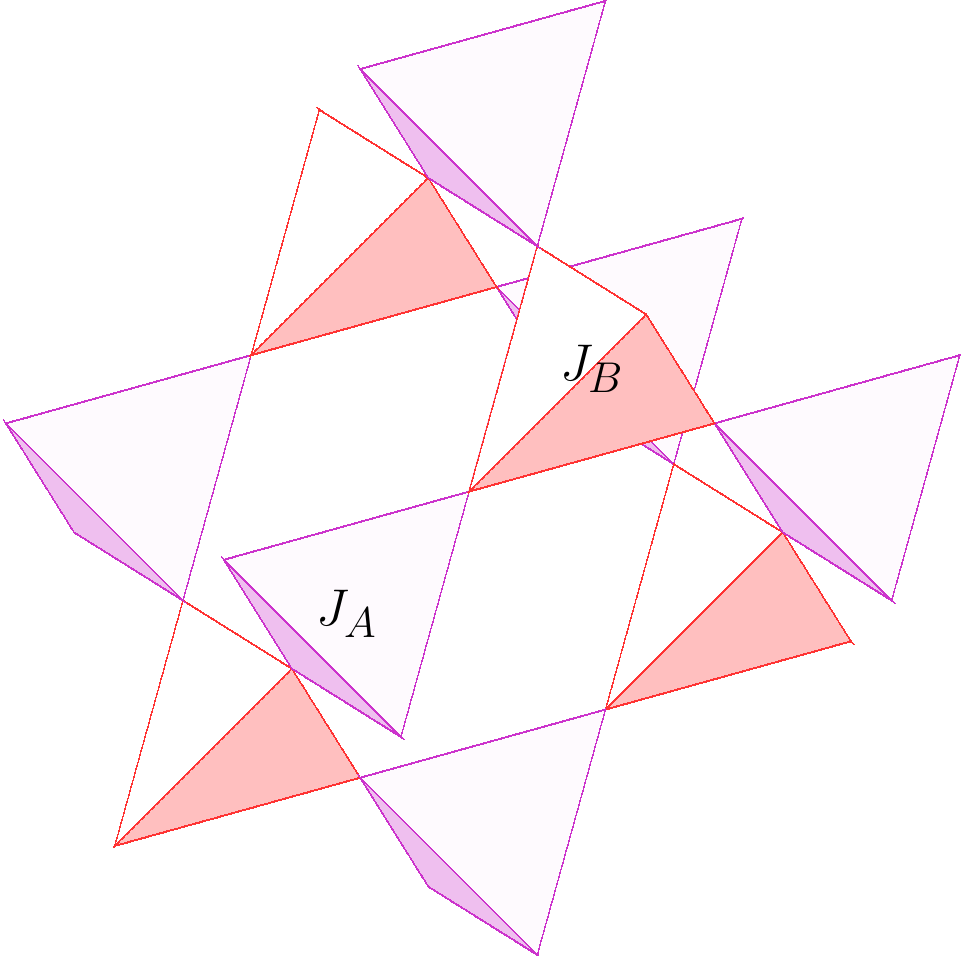}
\qquad\qquad
\includegraphics[width=4.cm]{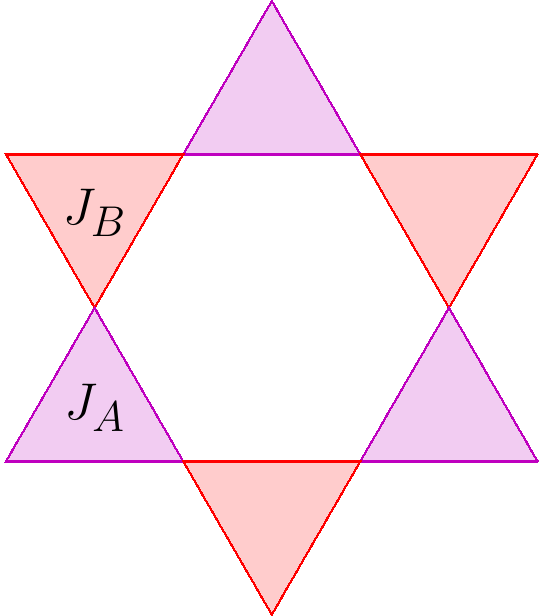}
\caption{Breathing pyrochlore (\textit{left}) and breathing kagome (\textit{right)}. The purple and red tetrahedra/triangles represent the A- and B-type of units with respectively $J_{A}$ and $J_{B}$ coupling constants. There are 4 spin sublattices in pryochlores and 3 spin sublattices in kagomes.}
\label{fig:latt}
\end{figure}
\section{Breathing lattices}

\subsection{Geometry}

Pyrochlores and kagomes are frustrated lattices made respectively of corner-sharing tetrahedra and triangles [Fig.~\ref{fig:latt}]. For convenience, tetrahedra and triangles shall be referred to as units. The centres of these units form bipartite lattices (respectively diamond and honeycomb). It is thus possible to define two types of units labeled by A and B [coloured in purple and red in Fig.~\ref{fig:latt}].\\

 Let $\mathbf{R}_{a}$ be the position of the centre of unit $a$. Within a A-type tetrahedron, the positions of the 4 spin sublattices with respect to $\mathbf{R}_{a}$ are
\begin{eqnarray}
\pmb{\delta}^{1}_{A}=\frac{1}{8} \begin{pmatrix} 1\\1\\1\end{pmatrix},\quad
\pmb{\delta}^{2}_{A}=\frac{1}{8} \begin{pmatrix} -1\\-1\\1\end{pmatrix},\quad
\pmb{\delta}^{3}_{A}=\frac{1}{8} \begin{pmatrix} -1\\1\\-1\end{pmatrix},\quad
\pmb{\delta}^{4}_{A}=\frac{1}{8} \begin{pmatrix} 1\\-1\\-1\end{pmatrix}.
\label{eq:sublpyro}
\end{eqnarray}
For a B-type tetrahedron, $\pmb{\delta}^{i}_{B}=-\pmb{\delta}^{i}_{A}$ for $i\in \{1,2,3,4\}$. All distances are given in units of the linear size of the cubic unit cell made of 16 pyrochlore sites.\\

Within a A-type triangle, there are 3 spin sublattices whose positions with respect to $\mathbf{R}_{a}$ are
\begin{eqnarray}
\pmb{\delta}^{1}_{A}=\frac{1}{4} \begin{pmatrix} -1\\-1/\sqrt{3}\end{pmatrix},\quad
\pmb{\delta}^{2}_{A}=\frac{1}{4} \begin{pmatrix} 1\\-1/\sqrt{3}\end{pmatrix},\quad
\pmb{\delta}^{3}_{A}=\frac{1}{4} \begin{pmatrix} 0\\2/\sqrt{3}\end{pmatrix}.
\label{eq:sublkag}
\end{eqnarray}
For a B-type triangle, $\pmb{\delta}^{i}_{B}=-\pmb{\delta}^{i}_{A}$ for $i\in \{1,2,3\}$. All distances are given in units of the linear size of the triangular unit cell made of 3 sites.

\subsection{Hamiltonian}

Let us consider a Hamiltonian with isotropic nearest-neighbour exchange
\begin{align}
\label{eq:ham}
\mathcal{H} = J_{A} \sum_{a \in A} \sum_{\langle ij \rangle} {\bf S}^{i}_{a} \cdot {\bf S}^{j}_{a} + J_{B} \sum_{b \in B} \sum_{\langle ij \rangle} {\bf S}^{i}_{b} \cdot {\bf S}^{j}_{b},
\end{align}
where $\mathbf{S}^{i}_{a}$ is a classical Heisenberg spin of unit length $|\mathbf{S}^{i}_{a}|=1$ on sublattice $i$ belonging to the unit $a$. The first and second terms of Eq.~(\ref{eq:ham}) run over all A- and B-type of units respectively. It will be useful to introduce the breathing factor $\alpha=J_{B}/J_{A}$~\cite{Okamoto2013a}.

\section{Mode spectrum of the interaction matrix}

\subsection{Interaction matrix}

In pyrochlore and kagome lattices, every spin belongs to two different units, indexed $a_{o}\in A$ and $b_{o}\in B$. This means that every spin can be written as $\mathbf{S}^{i}_{a_{o}}$ or as $\mathbf{S}^{i}_{b_{o}}$ with position $\mathbf{R}^{i}_{a_{o}}+\pmb{\delta}^{i}_{A}=\mathbf{R}^{i}_{b_{o}}+\pmb{\delta}^{i}_{B}=\mathbf{R}^{i}_{b_{o}}-\pmb{\delta}^{i}_{A}$. Keeping this in mind, one can define the Fourier transform of the spin degrees of freedom for a given sublattice $i$.
\begin{align}
\label{eq:TF}
\mathbf{\hat{S}}_{\bf q}^{i}
&=\frac{1}{\sqrt{N'}}\sum_{a\in A} \mathbf{S}_{a}^{i} \, {\rm e}^{-\imath {\bf q}.({\bf R}_a+\pmb{\delta}^{i}_{A})}
=\frac{1}{\sqrt{N'}}\sum_{b\in B} \mathbf{S}_{b}^{i} \, {\rm e}^{-\imath {\bf q}.({\bf R}_b+\pmb{\delta}^{i}_{B})},\\
\label{eq:FT}
\mathbf{S}_{a}^{i}
&=\frac{1}{\sqrt{N'}}\sum_{\bf q} \mathbf{\hat{S}}_{\bf q}^{i} \, {\rm e}^{+\imath {\bf q}.({\bf R}_a+\pmb{\delta}^{i}_{A})}
=\frac{1}{\sqrt{N'}}\sum_{\bf q} \mathbf{\hat{S}}_{\bf q}^{i} \, {\rm e}^{+\imath {\bf q}.({\bf R}_b+\pmb{\delta}^{i}_{B})}
= \mathbf{S}_{b}^{i},
\end{align}
where $N'$ is the total number of sites in a given spin sublattice. If $N$ is the total number of spins in the system and $s$ is the number of spin sublattices, then $N'=N/s$. The Fourier transform of Eq.~(\ref{eq:ham}) takes the form~\cite{Benton2015}
\begin{align}
\mathcal{H}
&=\frac{J_{A}}{2} \sum_{\bf q} \sum_{i=1}^{s} \sum_{\substack{j=1 \\ j\neq i}}^{s} \mathbf{\hat{S}}_{\bf q}^i \cdot \mathbf{\hat{S}}_{\bf -q}^j\; {\rm e}^{\imath {\bf q} . (\pmb{\delta}_{A}^{i}-\pmb{\delta}_{A}^{j})}
+ \frac{J_{B}}{2} \sum_{\bf q} \sum_{i=1}^{s} \sum_{\substack{j=1 \\ j\neq i}}^{s} \mathbf{\hat{S}}_{\bf q}^i \cdot \mathbf{\hat{S}}_{\bf -q}^j\; {\rm e}^{\imath {\bf q} . (\pmb{\delta}_{B}^{i}-\pmb{\delta}_{B}^{j})}\\
&=\frac{J_{A}}{2} \sum_{\bf q} \sum_{i=1}^{s} \sum_{\substack{j=1 \\ j\neq i}}^{s} \mathbf{\hat{S}}_{\bf q}^i \cdot \mathbf{\hat{S}}_{\bf -q}^j
\left({\rm e}^{\imath {\bf q} . (\pmb{\delta}_{A}^{i}-\pmb{\delta}_{A}^{j})} + \alpha\, {\rm e}^{-\imath {\bf q} . (\pmb{\delta}_{A}^{i}-\pmb{\delta}_{A}^{j})}\right).
\label{eq:H2}
\end{align}
The dispersion relations are obtained by diagonalizing the $\mathbf{q}-$dependent interaction matrix $\mathcal{K}_{\bf q}$ whose coefficients are $K_{\bf q}^{ii}=0$ and, for $i\neq j$,
\begin{eqnarray}
K_{\bf q}^{ij}={\rm e}^{\imath {\bf q} . (\pmb{\delta}_{A}^{i}-\pmb{\delta}_{A}^{j})} + \alpha\, {\rm e}^{-\imath {\bf q} . (\pmb{\delta}_{A}^{i}-\pmb{\delta}_{A}^{j})}
=(1+\alpha) \cos\left({\bf q} . (\pmb{\delta}_{A}^{i}-\pmb{\delta}_{A}^{j})\right)
+\imath (1-\alpha) \sin\left({\bf q} . (\pmb{\delta}_{A}^{i}-\pmb{\delta}_{A}^{j})\right).
\label{eq:Kij}
\end{eqnarray}
As opposed to the uniform pyrochlore~\cite{Reimers1991,Gingras2001} and kagome~\cite{Garanin99a} lattices ($\alpha=1$), the coefficients of $\mathcal{K}_{\bf q}$ are intrinsically complex for breathing lattices~\cite{Benton2015}. However since $\mathcal{K}_{\bf q}$ is hermitian, the eigenvalues remain of course real. For most of the paper, the discussion will be focused on $J_{A}>0$ and $-1\leq\alpha\leq 1$, because the mode spectrum for $J_{A}<0$ and $|\alpha|> 1$ can be easily derived by a rescaling of the coupling constants.

\subsubsection{Breathing pyrochlore}

The four eigenvalues of $\mathcal{K}_{\mathbf{q}}$ are given by
\begin{align}
\left\{\begin{array}{ll}
\lambda^{1}_{\bf q} &= \alpha + 1 - 2 \sqrt{\alpha ^2-\alpha +\alpha  \cos \left(\frac{q_x}{2}\right) (\cos \left(\frac{q_y}{2}\right)+\cos \left(\frac{q_z}{2}\right))+\alpha  \cos \left(\frac{q_y}{2}\right) \cos \left(\frac{q_z}{2}\right)+1} \\
\lambda^{2}_{\bf q} &= \alpha + 1 +2 \sqrt{\alpha ^2-\alpha +\alpha  \cos \left(\frac{q_x}{2}\right) (\cos \left(\frac{q_y}{2}\right)+\cos \left(\frac{q_z}{2}\right))+\alpha  \cos \left(\frac{q_y}{2}\right) \cos \left(\frac{q_z}{2}\right)+1}\\
\lambda^{3}_{\bf q} &= -\alpha -1 \\
\lambda^{4}_{\bf q} &= -\alpha -1
\end{array}\right.
\label{eq:lambdapyro}
\end{align}	
with two degenerate flat bands, as explained in Sec.~\ref{sec:TB}. 

\subsubsection{Breathing kagome}

The three eigenvalues of $\mathcal{K}_{\mathbf{q}}$ are given by
\begin{align}
\left\{\begin{array}{ll}
\lambda^{1}_{\bf q} &= \dfrac{1}{2} \left(1+ \alpha -\sqrt{9 \alpha ^2-6 \alpha +8 \alpha  \cos (q_x)+16 \alpha  \cos (\frac{q_x}{2}) \cos \left(\frac{\sqrt{3}}{2}
		q_y\right)+9}\right) \\
\lambda^{2}_{\bf q} &= \dfrac{1}{2} \left(1 +\alpha +\sqrt{9 \alpha ^2-6 \alpha +8 \alpha  \cos (q_x)+16 \alpha  \cos (\frac{q_x}{2}) \cos \left(\frac{\sqrt{3}}{2}
		q_y\right)+9}\right)\\
\lambda^{3}_{\bf q} &= -\alpha -1
\end{array}\right.
\label{eq:lambdakag}
\end{align}
with one flat band, as explained in Sec.~\ref{sec:TB}. 

\subsection{Origin of the flat bands}
\label{sec:TB}
\begin{figure}[h]
\centering
\includegraphics[width=15cm]{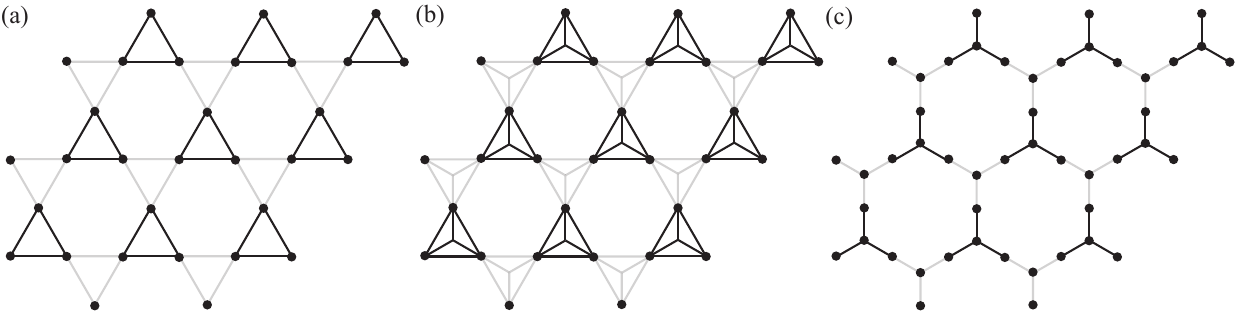}
\caption{The process of making an auxiliary hybrid lattice from a breathing kagome lattice. (a) A breathing kagome lattice. (b) Sites are added in the center of triangle units. (c) The bonds of original breathing kagome lattice is removed.}
\label{fig:kagome_to_honeycomb}
\end{figure}

In breathing pyrochlore and kagome lattices, some of the bands are completely flat, as shown in (\ref{eq:lambdapyro}) and (\ref{eq:lambdakag}).
The origin of these flat bands can be clarified by the factorization of the Hamiltonian (\ref{eq:ham}), as was done for the tight-binding model on complete graphs~\cite{Katsura2010}.
To this aim, we rewrite the Hamiltonian (\ref{eq:ham}) as
\begin{eqnarray}
\mathcal{H} = \frac{1}{2}\,\sum_{m,n}\hat{H}_{mn}\,{\mathbf S}_m\cdot{\mathbf S}_{n}.
\label{eq:ham2}
\end{eqnarray}
Here, ${\mathbf S}_m$ is the spin defined at the site, $m$, and the summation with $m$ and $n$ is made over the entire lattice.
The coefficient matrix, $\hat{H}$, can be factorized by introducing an auxiliary hybrid lattice, as we show in Fig.~\ref{fig:kagome_to_honeycomb} for the breathing kagome lattice.  
Firstly, we put sites at the centers of units, and secondly, we connect them to the original sites [Fig.~\ref{fig:kagome_to_honeycomb} (b)], and finally we remove the original bonds [Fig.~\ref{fig:kagome_to_honeycomb} (c)].
The resultant lattice is the decorated diamond (honeycomb) lattice for breathing pyrochlore (kagome) lattice.

The new auxiliary lattice consists of two classes of sites: those of the original lattice, and those at the centres of the units of the original lattice. We call the former class $\ell$, and the latter, $c$. The number of sites in class $\ell$ and $c$ are defined as $N_\ell$ and $N_c$, respectively. These are given as $N_\ell = N$ and $N_c = 2N'=2N/s$ for both lattices.

On this auxiliary lattice, we introduce two rectangular matrices, $\hat{H}^{c\leftarrow \ell}$ and $\hat{H}^{\ell\leftarrow c} = (\hat{H}^{c\leftarrow \ell})^T$. $\hat{H}^{c\leftarrow \ell}$ is a $N_c\times N_\ell$ incidence matrix, and its column (row) indices are related to the sites in $\ell$ ($c$). We set its $(m, n)$ component, $(\hat{H}^{c\leftarrow \ell})_{mn}$, to be $1$, if the corresponding sites, $m\in c$ and $n\in \ell$ are connected, and to be $0$, otherwise. Being a $N_c\times N_\ell$ rectangular matrix, the rank of $\hat{H}^{c\leftarrow \ell}$ is bounded by $\min(N_{c},N_{\ell})=N_{c}$\footnote{It is actually possible to prove that rank $\hat{H}^{c\leftarrow \ell}=N_{c}-1$ by counting the number of localized modes~\cite{bergman08a}.}. Given these definitions, the matrix $\hat{H}$ of Eq.~(\ref{eq:ham2}) can be factorized as
\begin{eqnarray}
\hat{H} = J_A\hat{H}^{\ell\leftarrow c}\,\hat{\tau}\,\hat{H}^{c\leftarrow \ell}
- J_A(1 + \alpha)\, \mathbb{1}_{\ell},
\end{eqnarray}
where $\mathbb{1}_{\ell}$ is the $N_\ell\times N_\ell$ identity matrix and $\hat{\tau}$ is a $N_c\times N_c$ diagonal matrix, $\hat{\tau}_{mn} = \tau_m\delta_{mn}$, with $\tau_m = 1 (\alpha)$ if $m$ belongs to sublattice A (B).

This expression immediately leads to the protection of the flat bands present in the uniform lattices. Firstly, suppose there is a state annihilated by $\hat{H}^{c\leftarrow \ell}$, then this state serves as an eigenstate of $\hat{H}$ with eigenvalue, $-J_A(1 + \alpha)$, and is independent of $\alpha$. Secondly, there are in fact at least $(N_\ell-N_c)\,=(s-2)N'$ such states, since $\hat{H}^{c\leftarrow \ell}$ acts on a space of dimension $N_\ell$, while the rank of $\hat{H}^{c\leftarrow \ell}$ is bounded by $N_c$. Consequently, $\hat{H}$ has at least $(s-2)N'$ degenerate states with eigenvalue, $-J_A(1 + \alpha)$. This argument proves that the $s-2=2\, (1)$ flat bands of the uniform pyrochlore (kagome) lattices see their energy shifted by the breathing anisotropy, but are not destroyed and conserves their degeneracy. In addition, the presence of new flat bands is allowed, which is precisely what happens for $\alpha=0$ (see below).

\begin{figure}[b]
\centering
\includegraphics[width=7.5cm]{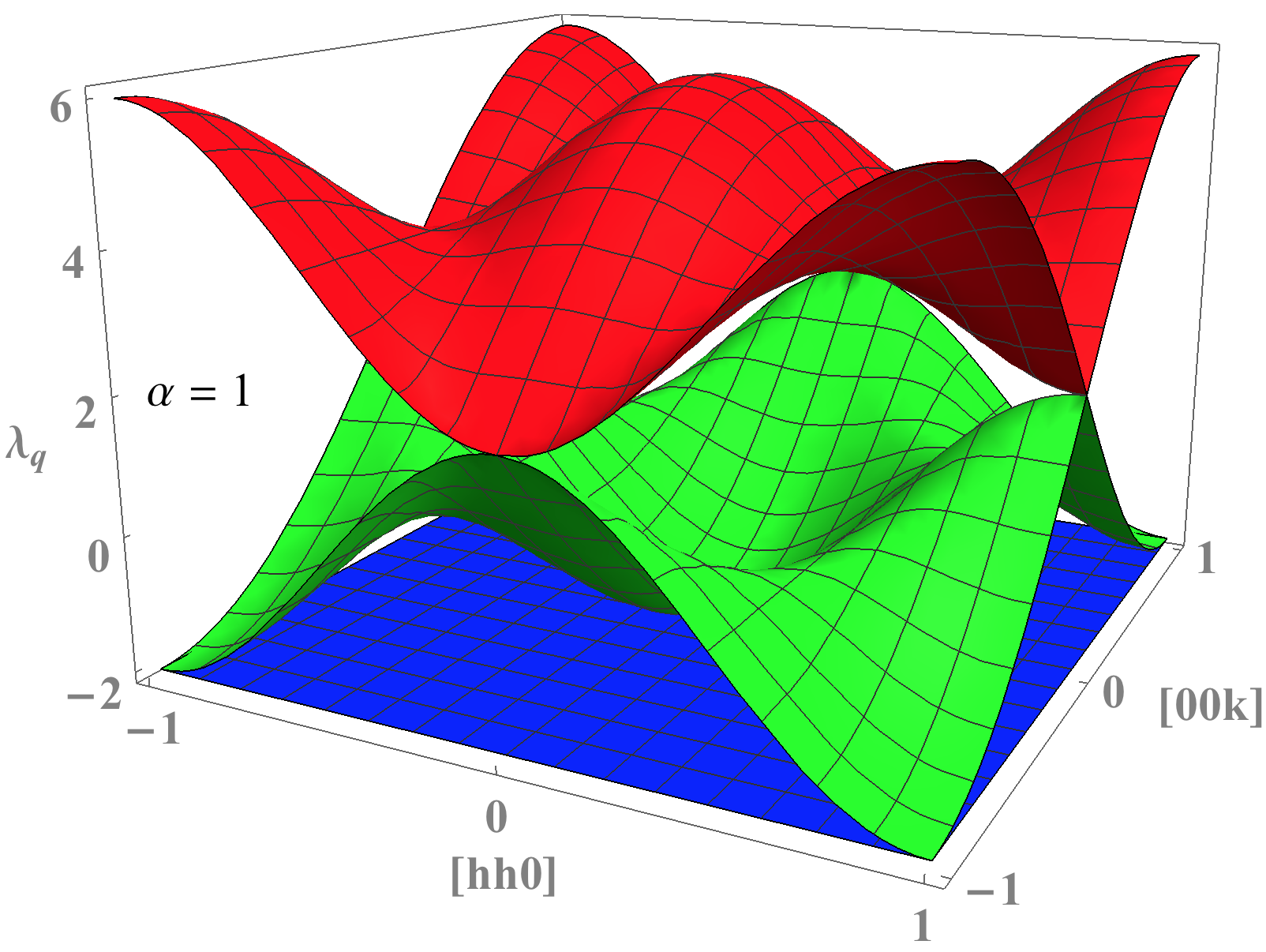}
\includegraphics[width=7.5cm]{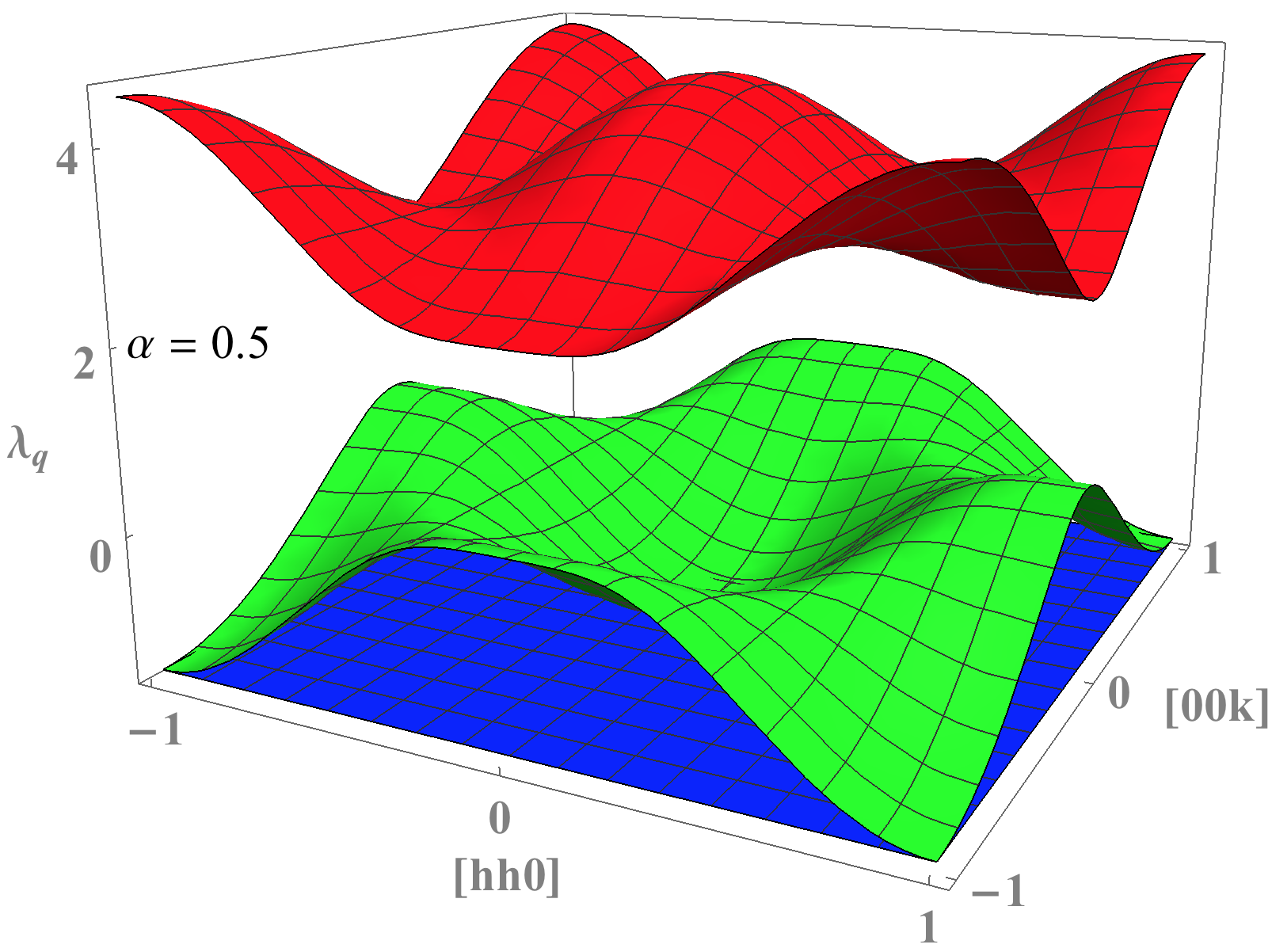}
\caption{Mode spectrum of the interaction matrix for breathing pyrochlores in the [hhk] plane at $\alpha=1$ (see Ref.~\cite{Reimers1991}) and $\alpha=0.5$. The ground state is a doubly degenerate flat band with gapless excitations. A gap between the dispersive bands opens for $|\alpha|\neq 1$. Wavevectors are in $2\pi$ units.}
\label{fig:pyrop}
\end{figure}
\begin{figure}[b]
\centering
\includegraphics[width=7.5cm]{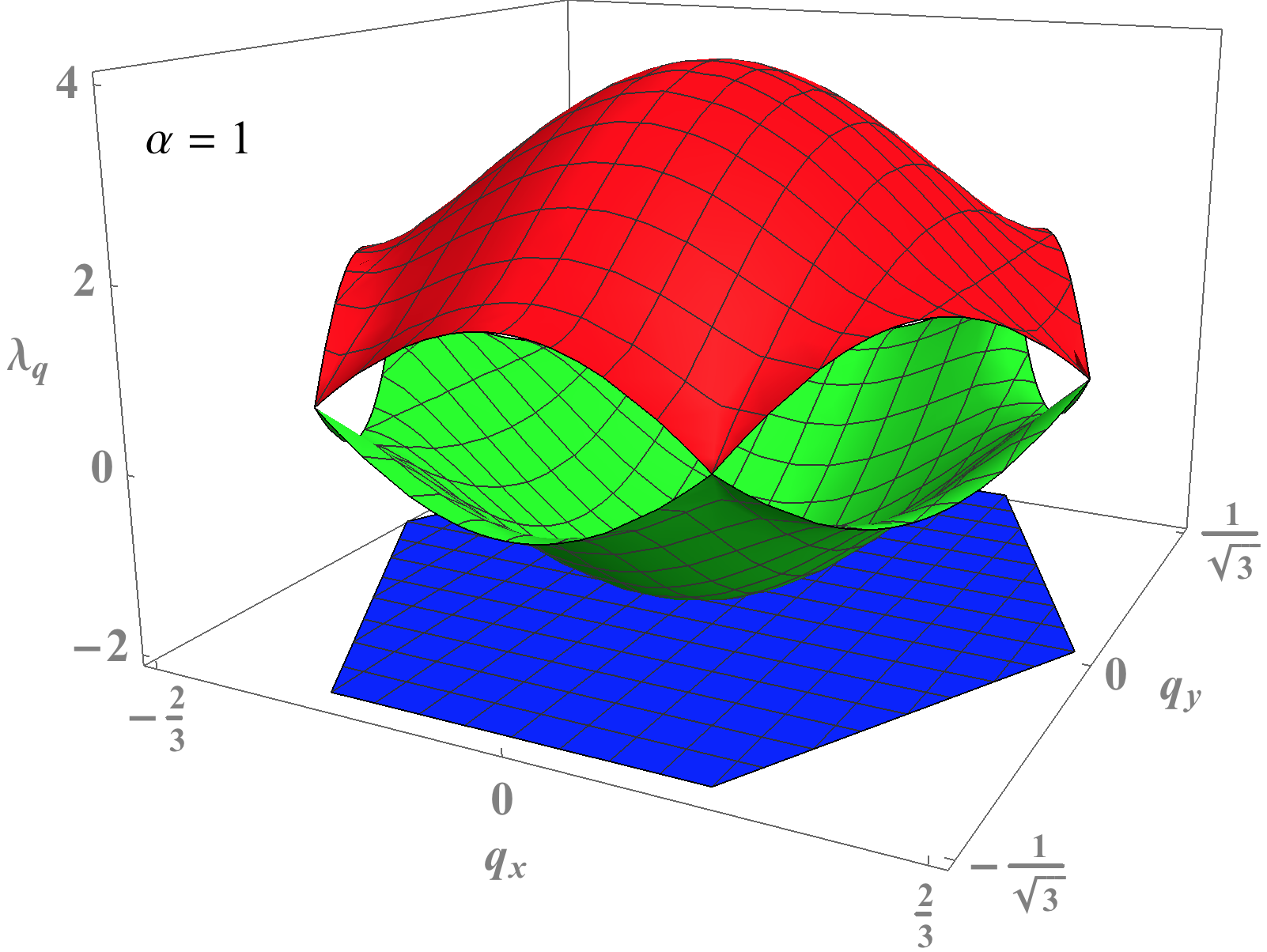}
\includegraphics[width=7.5cm]{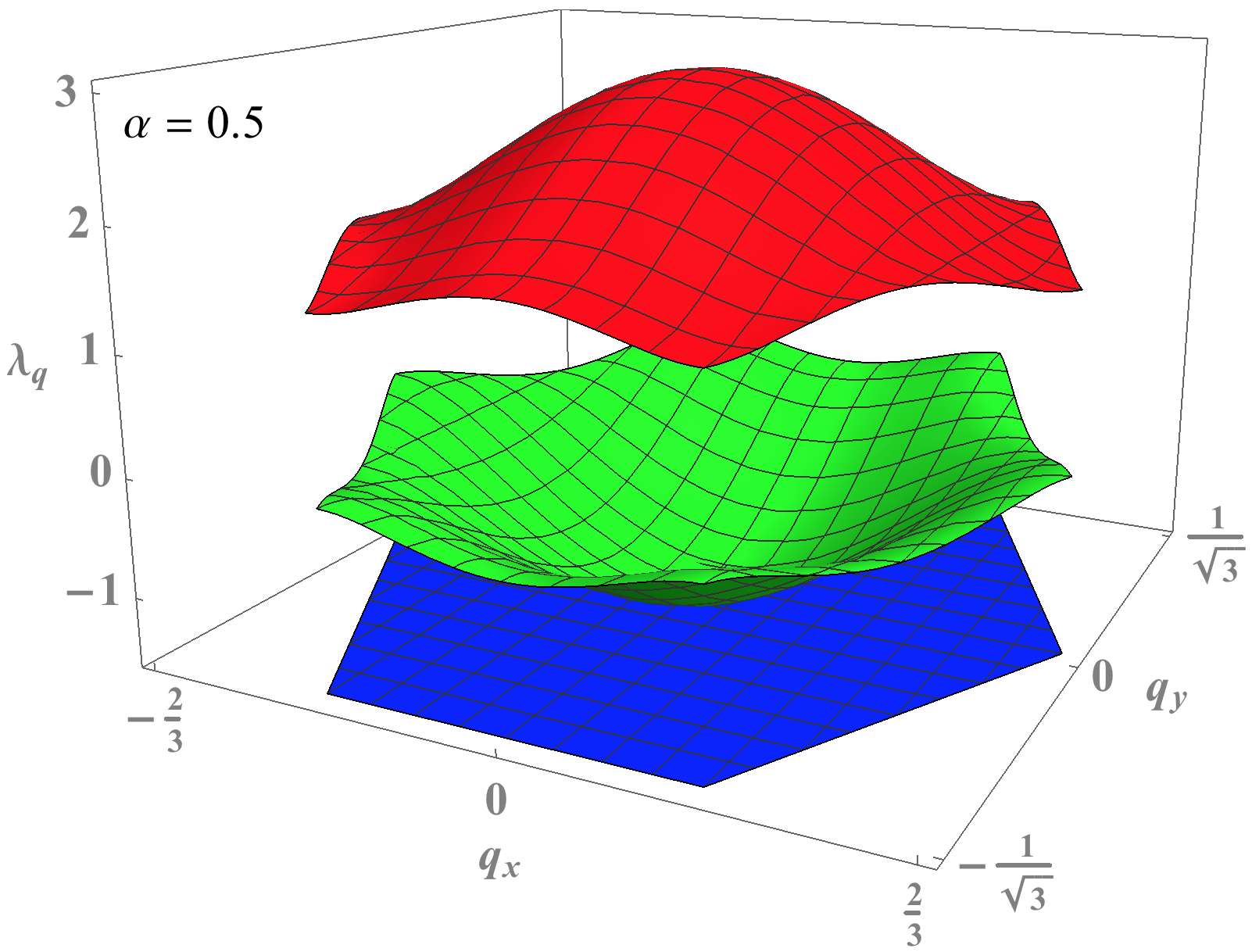}
\caption{Mode spectrum of the interaction matrix for breathing kagomes at $\alpha=1$ (see Ref.~\cite{Garanin99a}) and $\alpha=0.5$. The ground state is a unique flat band with gapless excitations. A gap between the dispersive bands opens for $|\alpha|\neq 1$. Wavevectors are in $2\pi$ units.}
\label{fig:kagp}
\end{figure}
\subsection{Gap opening when $|J_{A}|\neq|J_{B}|$}

The mode spectra for regular pyrochlore and kagome lattices are known to be gapless~\cite{Reimers1991,Garanin99a}. Here we find that for $\alpha>0$ excitations out of the flat bands remain gapless, but the breathing anisotropy opens a gap between the dispersive bands $\lambda^{1}_{\bf q}$ and $\lambda^{2}_{\bf q}$ when $|\alpha|\neq 1$ [Figs.~\ref{fig:pyrop} \& \ref{fig:kagp}]. From Eqs.~(\ref{eq:lambdapyro}) and (\ref{eq:lambdakag}), one gets
\begin{eqnarray}
\Delta_{\rm pyrochlore}=4\left|1-|\alpha|\right|, \qquad \Delta_{\rm kagome}=3\left|1-|\alpha|\right|.
\label{eq:gapp}
\end{eqnarray}

Starting from $\alpha=1$ and decreasing its value, $\lambda^{1}_{\bf q}$ and $\lambda^{2}_{\bf q}$ become less and less dispersive [Figs.~\ref{fig:pyrop} \& \ref{fig:kagp}] and flat at $\alpha=0$. For decoupled antiferromagnetic units, the ground state is made of three (resp. two) degenerate flat bands in pyrochlores (resp. kagomes).

\begin{figure}[b]
\centering
\includegraphics[width=7.5cm]{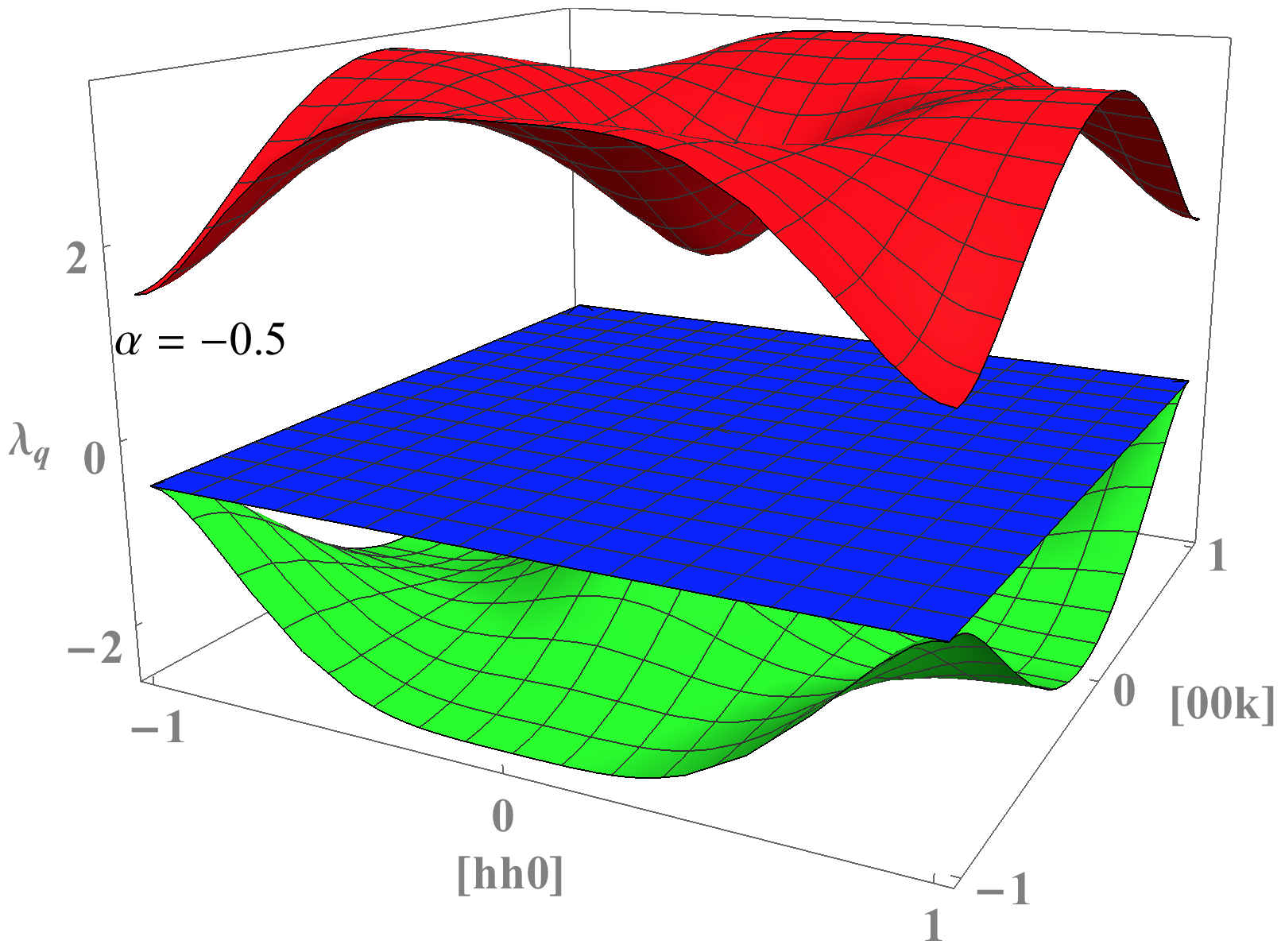}
\includegraphics[width=7.5cm]{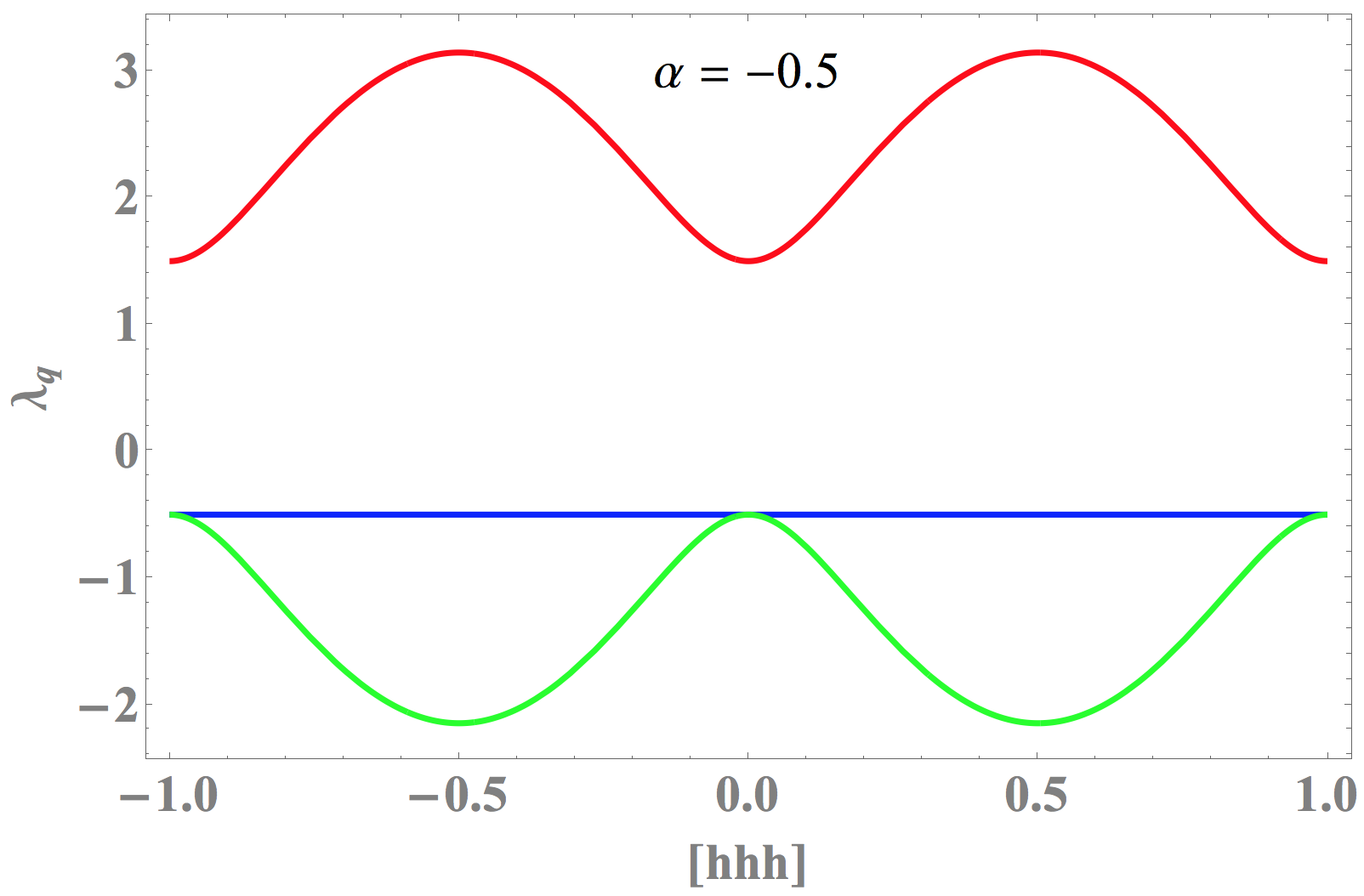}\\
\includegraphics[width=7.5cm]{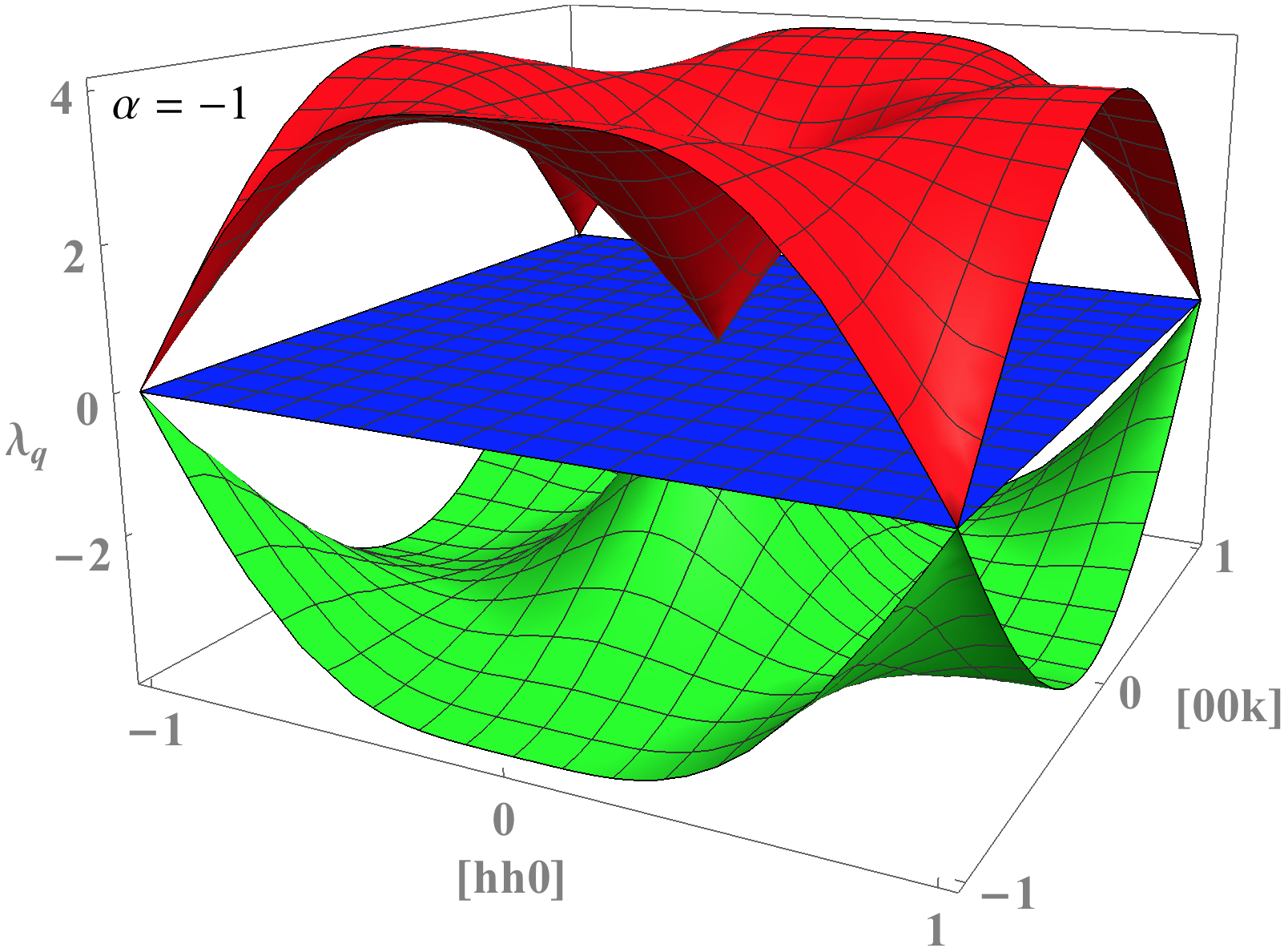}
\includegraphics[width=7.5cm]{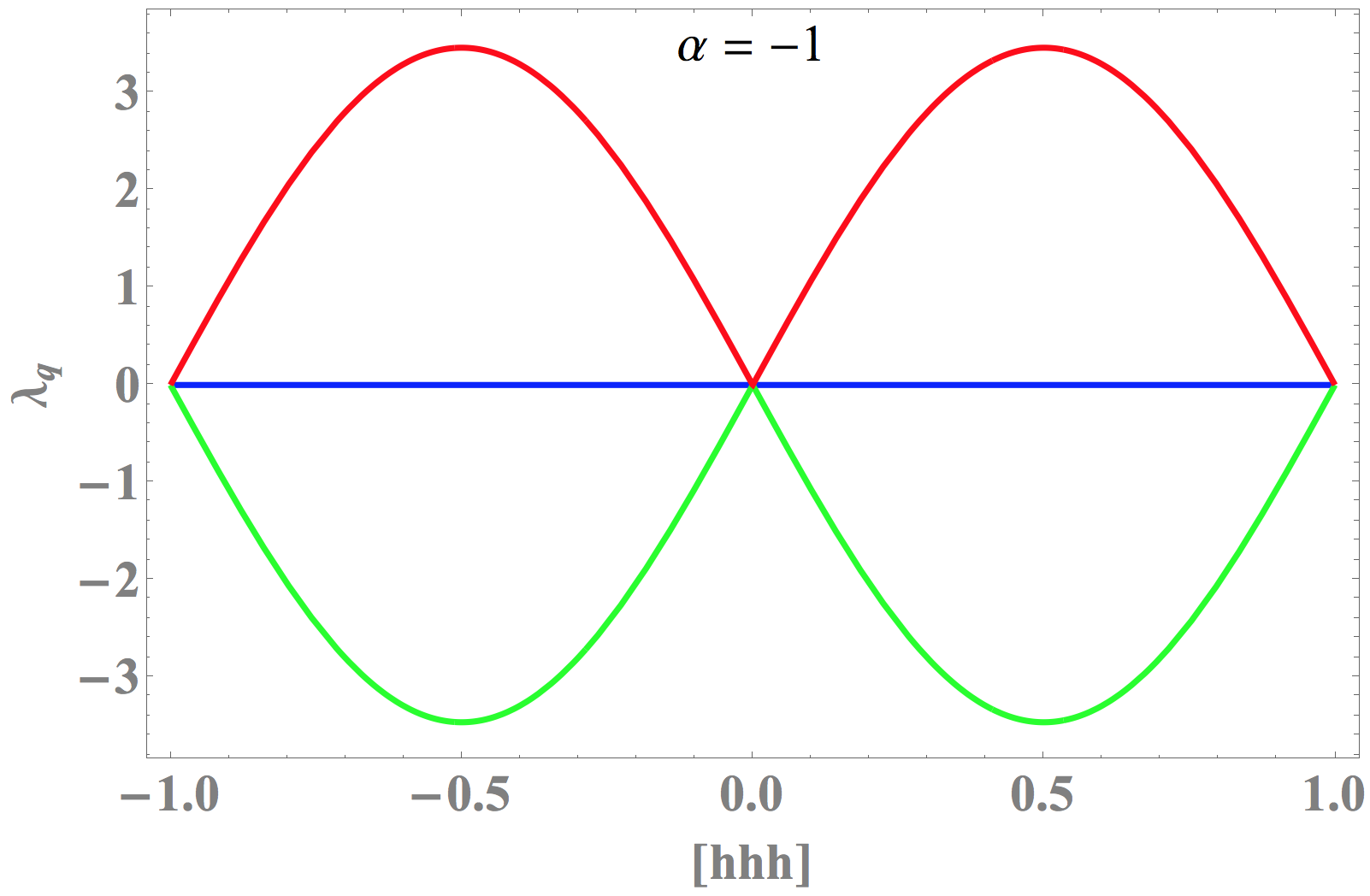}\\
\includegraphics[width=7.5cm]{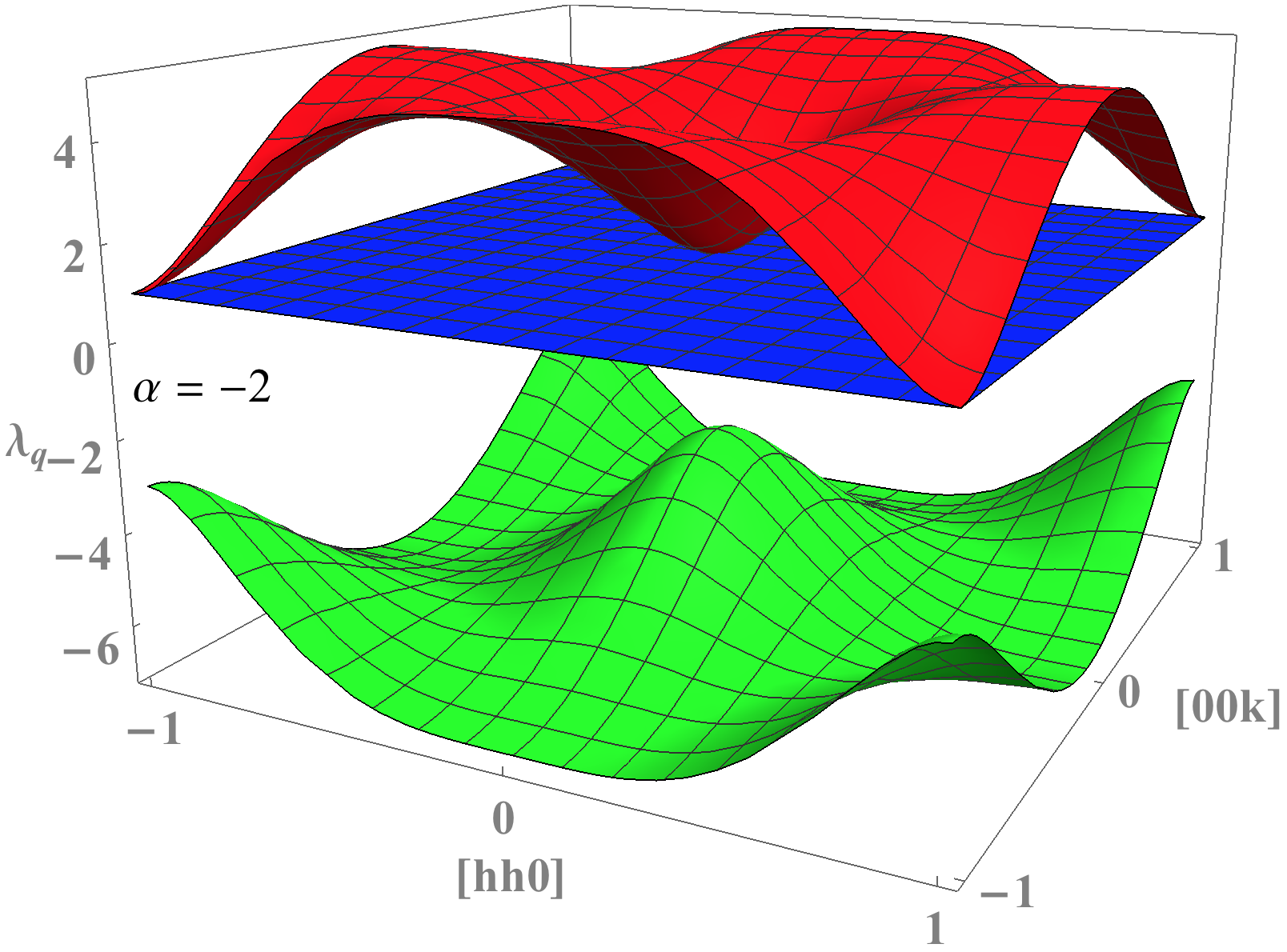}
\includegraphics[width=7.5cm]{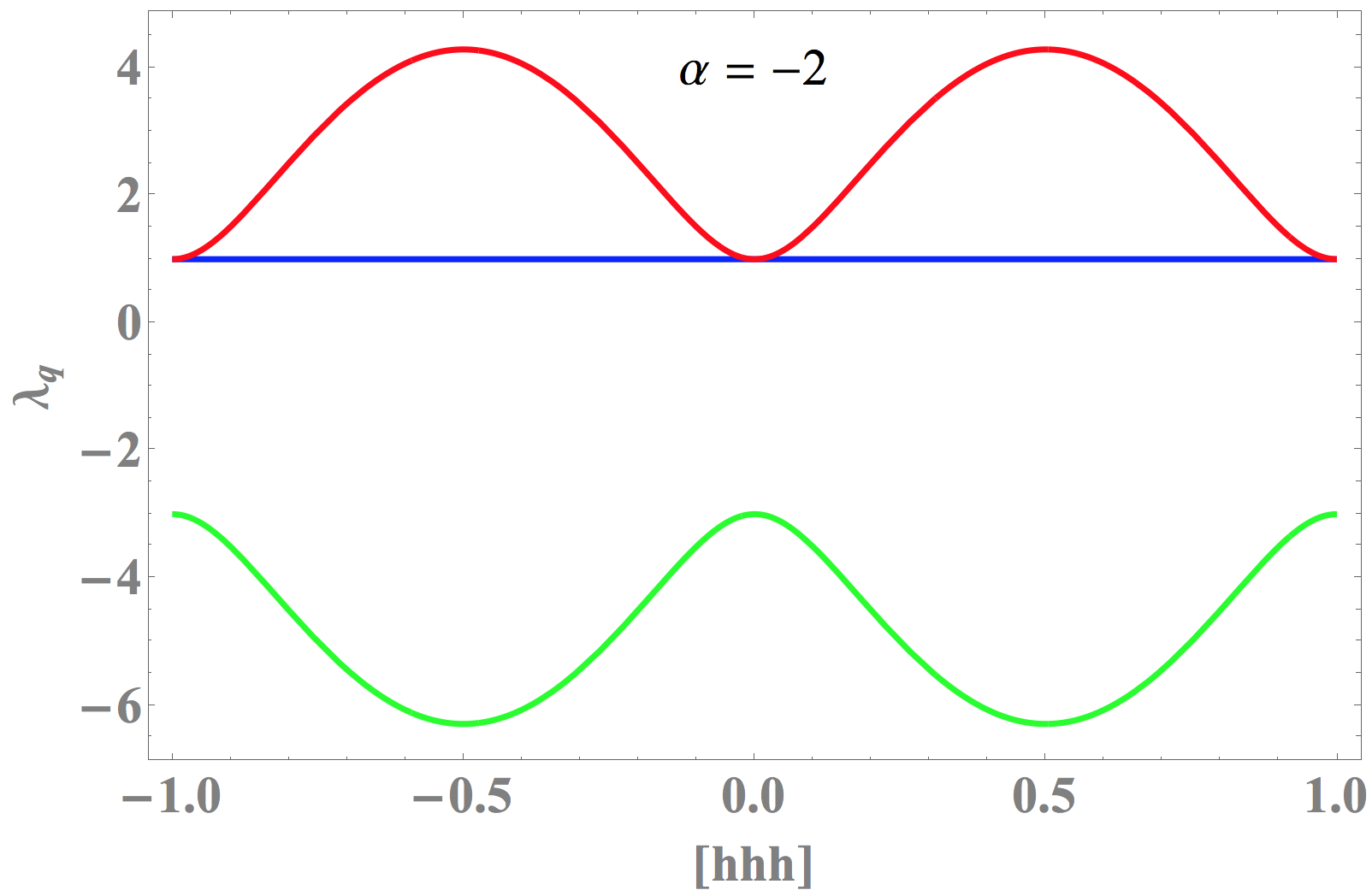}
\caption{Mode spectrum of $\mathcal{K}_{\mathbf{q}}$ for breathing pyrochlores and $\alpha\in\{-0.5, -1,-2\}$. The left column shows the entire [hhk] plane, while the right column shows a cut along the [hhh] line. For $\alpha<0$, the flat bands are not part of the ground state anymore. The gap closes at $\alpha=-1$ where a Dirac cone appears at the $\Gamma$ point. Wavevectors are in $2\pi$ units.}
\label{fig:pyrom}
\end{figure}
\begin{figure}[t]
\centering
\includegraphics[width=7.5cm]{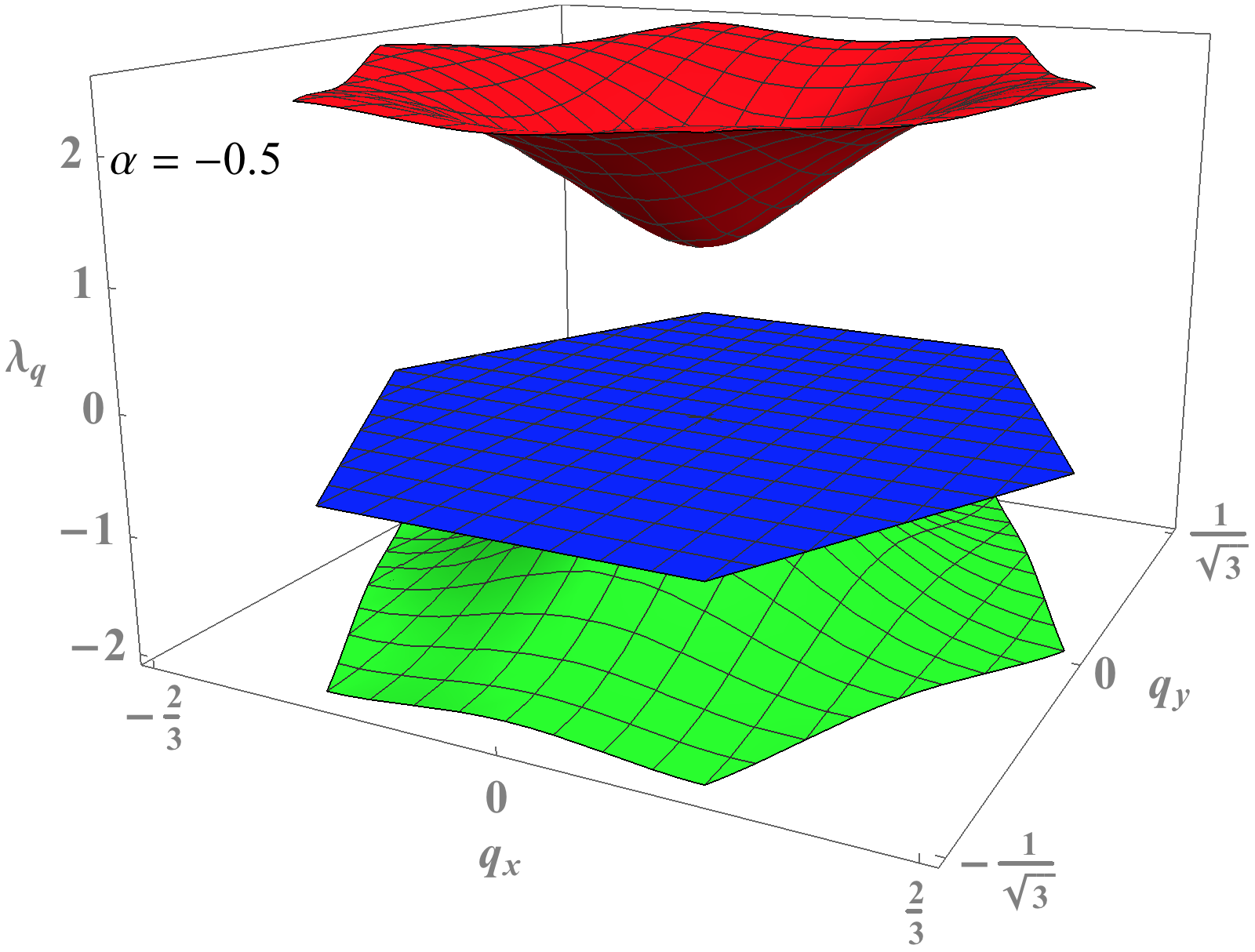}
\includegraphics[width=7.5cm]{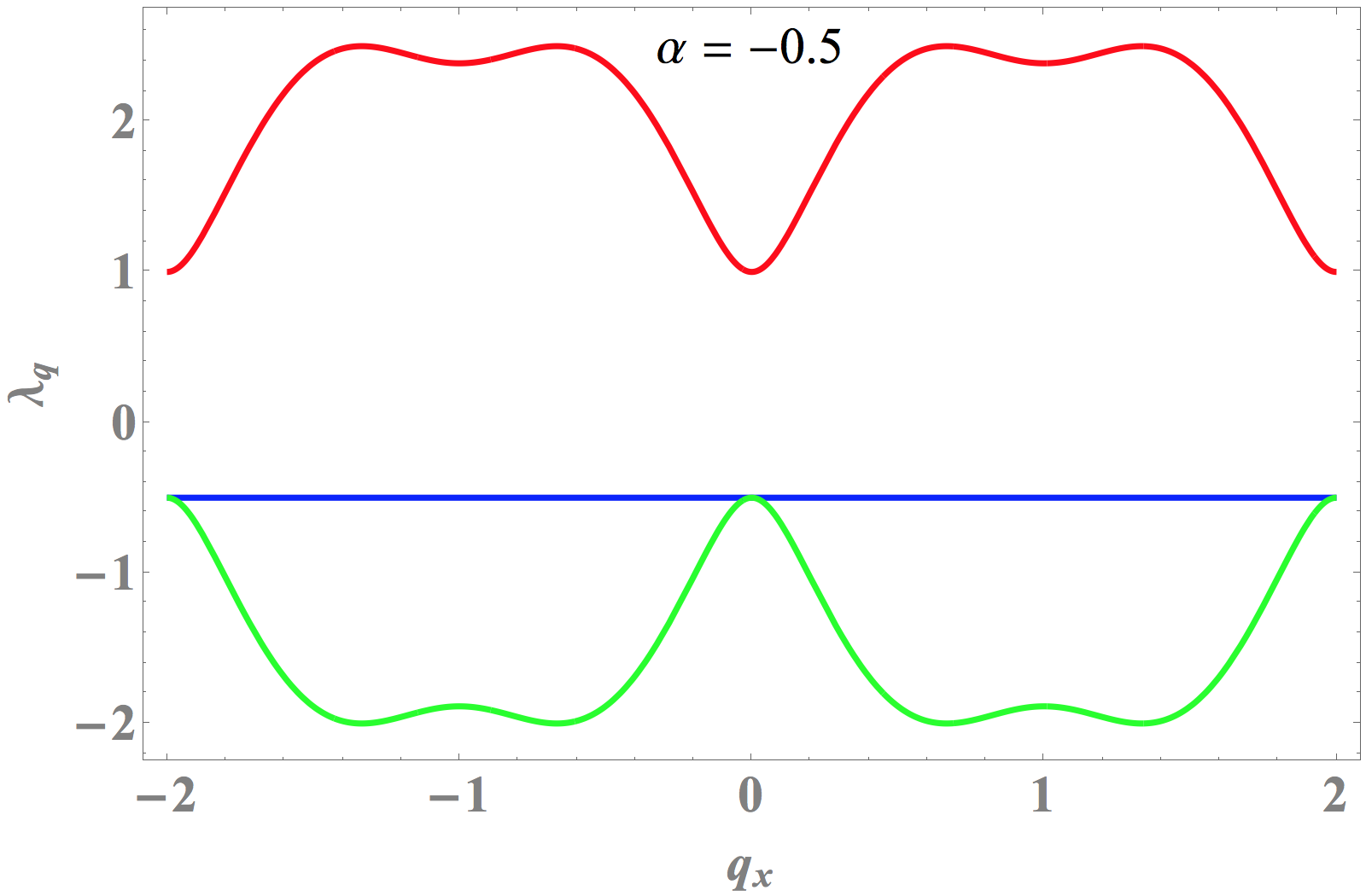}\\
\includegraphics[width=7.5cm]{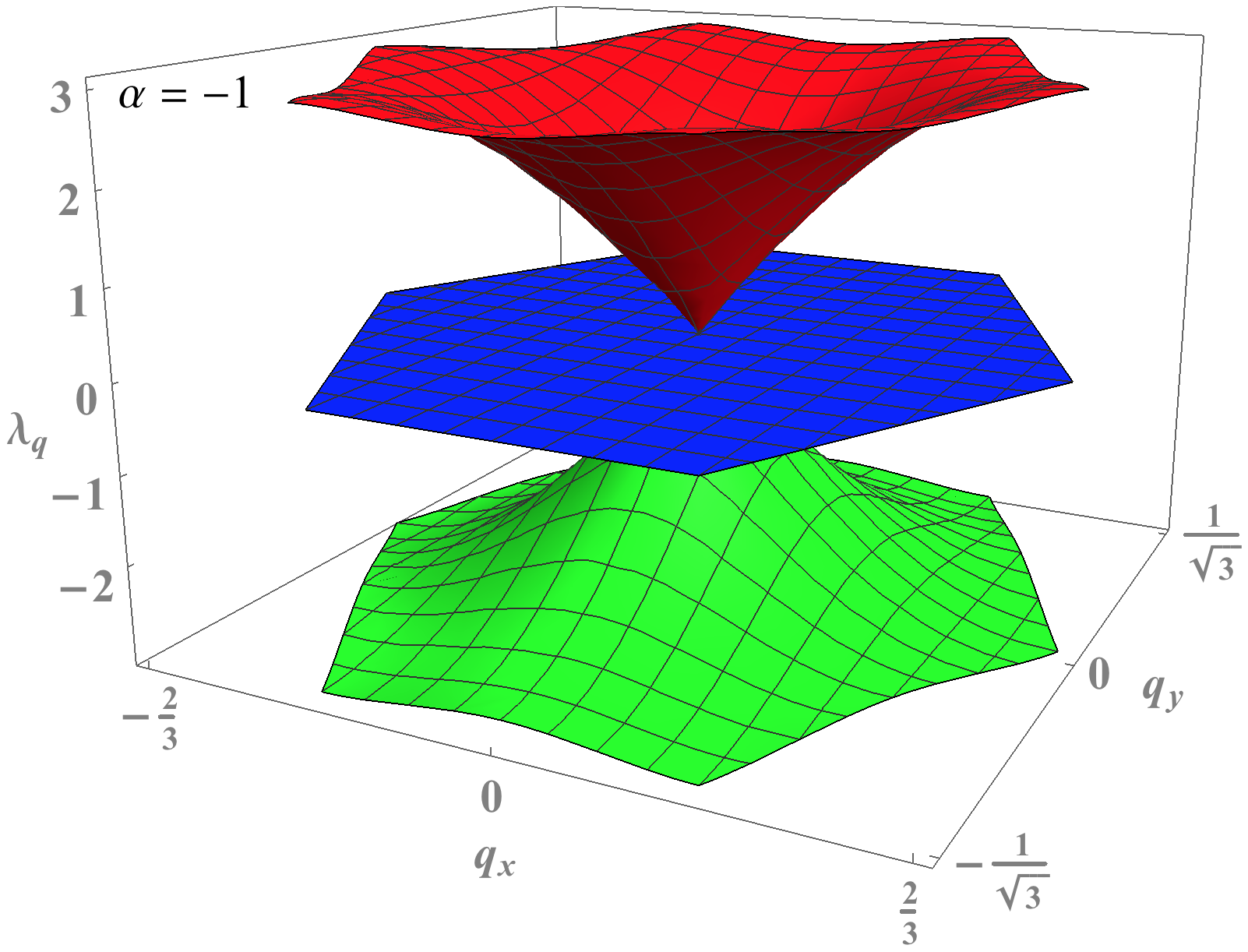}
\includegraphics[width=7.5cm]{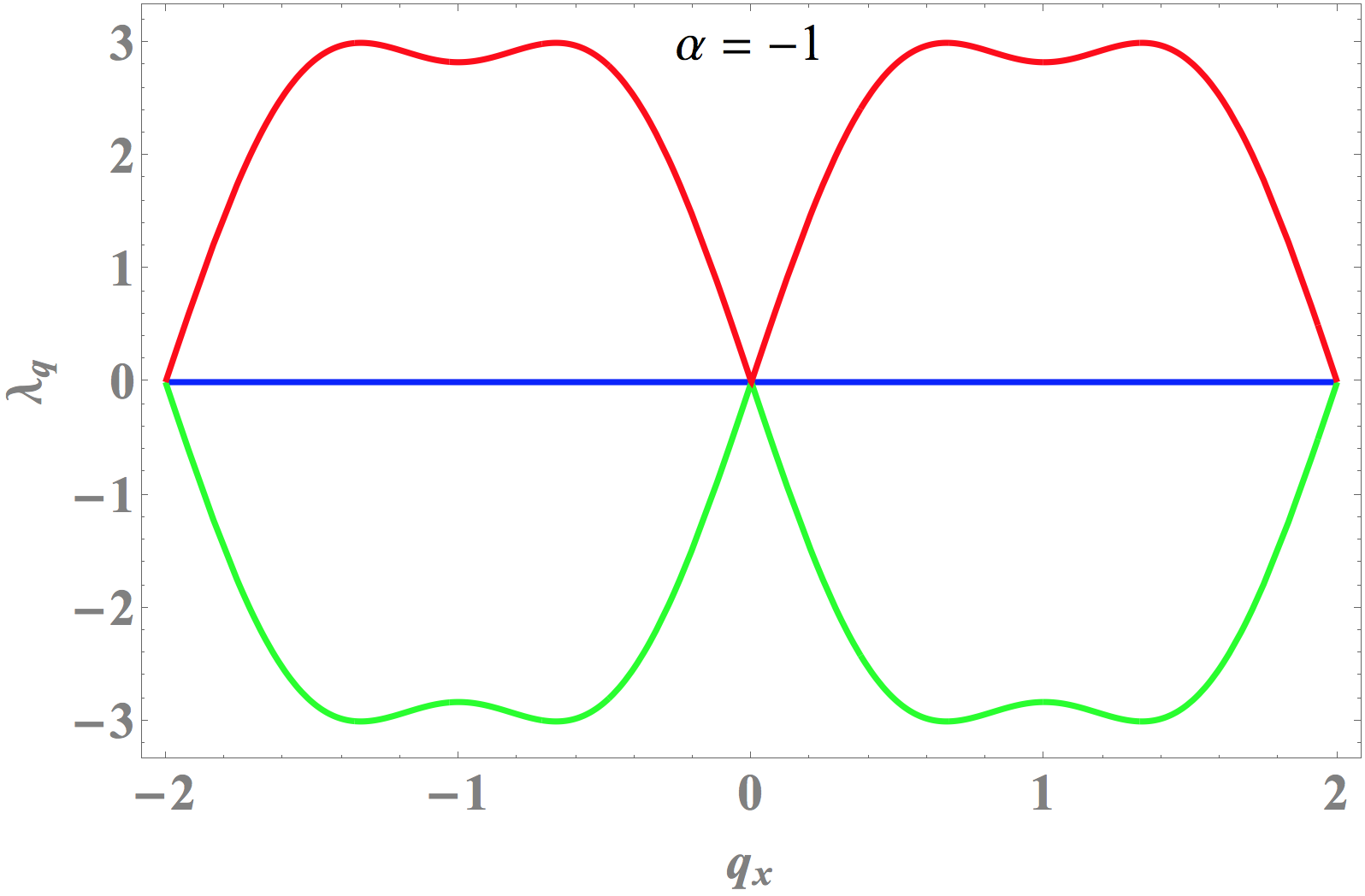}\\
\includegraphics[width=7.5cm]{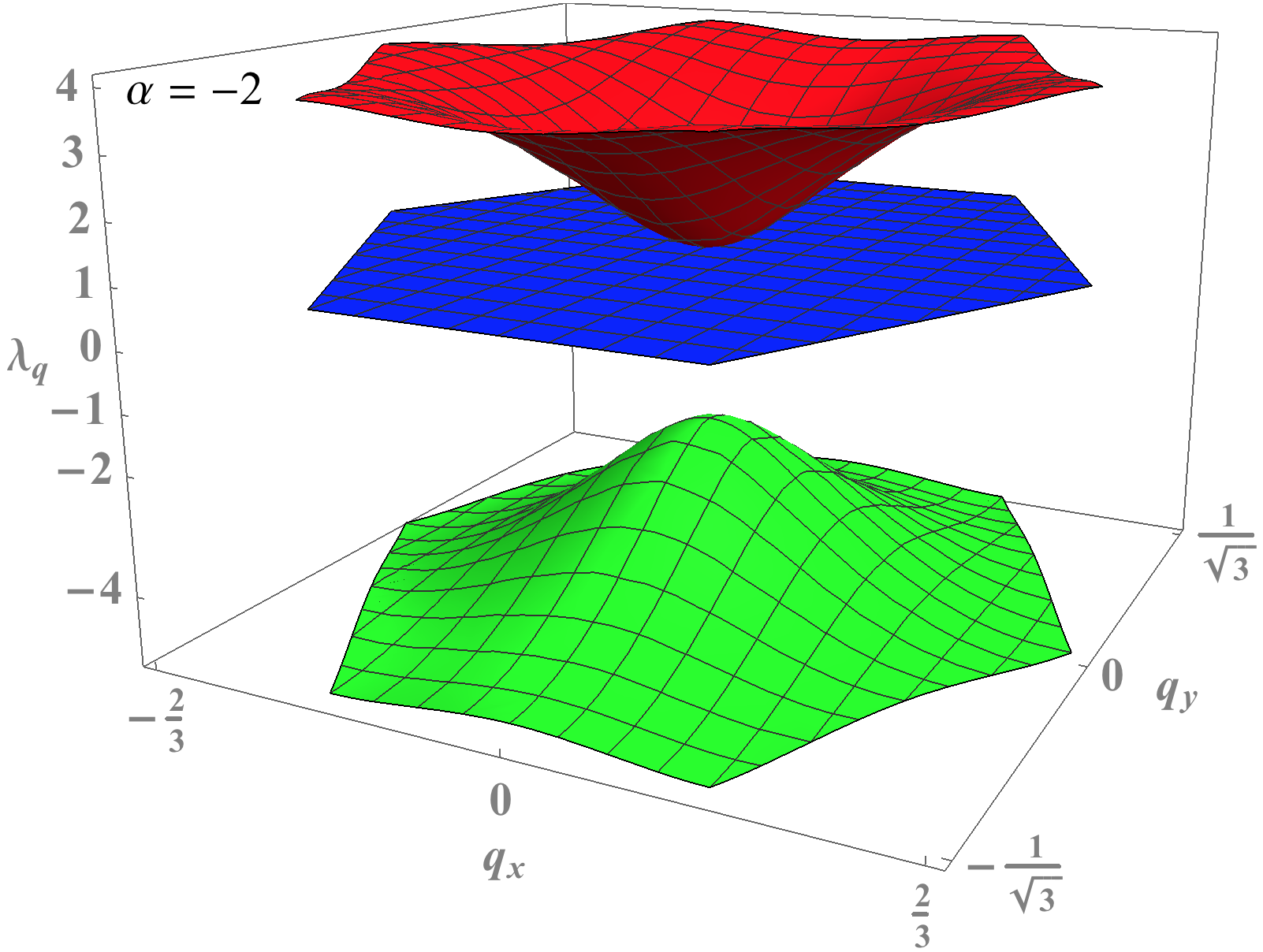}
\includegraphics[width=7.5cm]{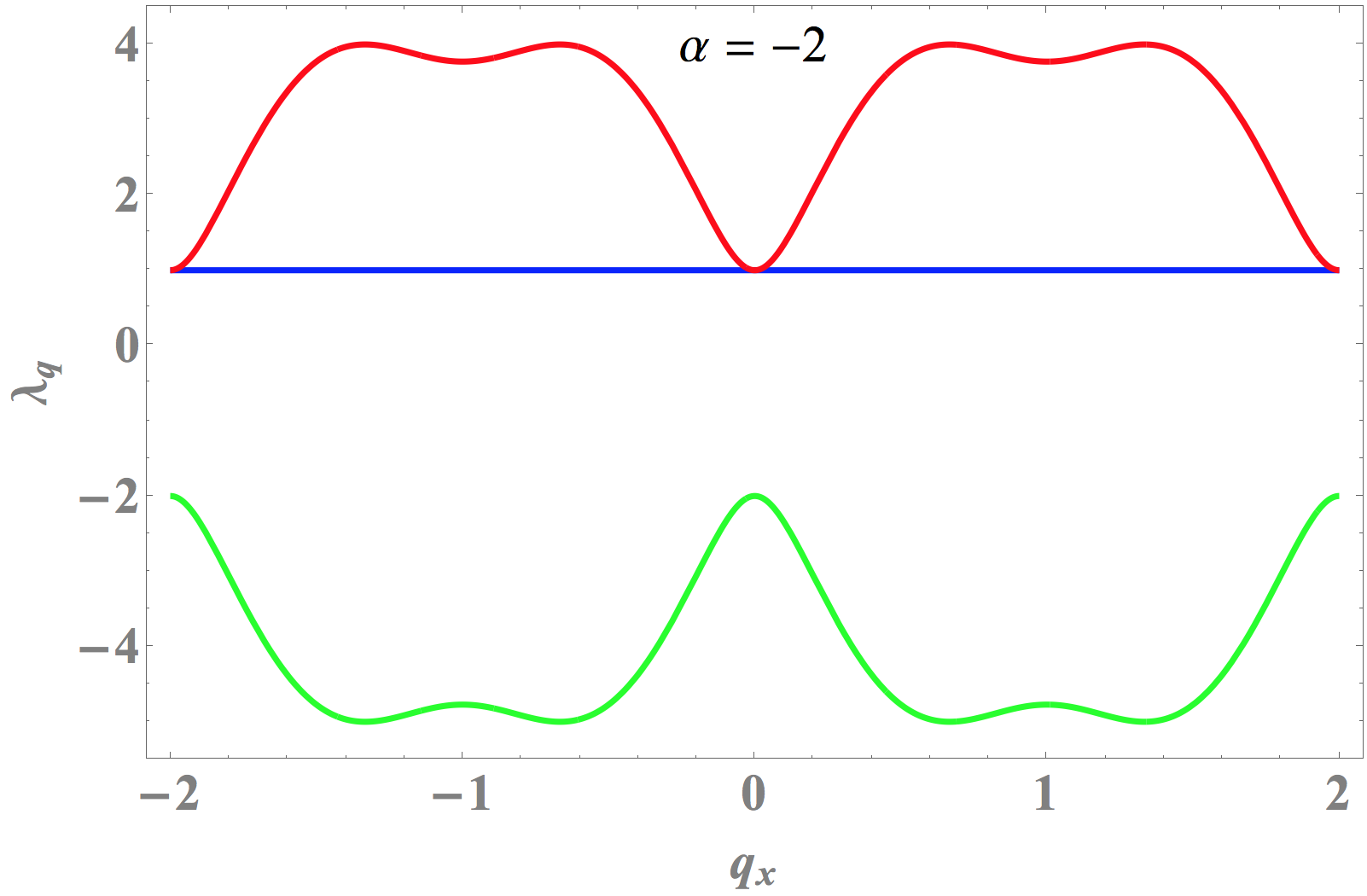}
\caption{Mode spectrum of $\mathcal{K}_{\mathbf{q}}$ for breathing kagomes and $\alpha\in\{-0.5, -1,-2\}$. The left column shows the entire ($q_{x},q_{y}$) plane, while the right column shows a cut along the $q_{y}=0$ line. For $\alpha<0$, the flat bands are not part of the ground state anymore. The gap closes at $\alpha=-1$ where a Dirac cone appears at the $\Gamma$ point. Wavevectors are in $2\pi$ units.}
\label{fig:kagm}
\end{figure}
\subsection{Particle-hole symmetry and Dirac cones for $J_{A}=-J_{B}$}

The breathing factor $\alpha$ can be seen as a way to tune the relative importance of the real and imaginary parts in $\mathcal{K}_{\mathbf{q}}$ [see Eq.~(\ref{eq:Kij})]. When $\alpha=-1$, $\mathcal{K}_{\mathbf{q}}$ becomes purely imaginary, $\mathcal{K}_{\mathbf{q}}=\imath \Omega_{\mathbf{q}}$ with $\Omega_{\mathbf{q}}\in \mathcal{M}(s,\mathbb{R})$. Since $\mathcal{K}_{\mathbf{q}}$ is hermitian, its eigenvalues are real. Let $\mathbf{V}_{\mathbf{q}}$ be an eigenvector of $\mathcal{K}_{\mathbf{q}}$ with eigenvalue $\lambda_{\mathbf{q}}$,
\begin{eqnarray}
\mathcal{K}_{\mathbf{q}} \mathbf{V}_{\mathbf{q}}= \imath \Omega_{\mathbf{q}} \mathbf{V}_{\mathbf{q}} = \lambda_{\mathbf{q}} \mathbf{V}_{\mathbf{q}}.
\label{eq:eigenv}
\end{eqnarray}
The complex conjugate of Eq.~(\ref{eq:eigenv}) gives
\begin{eqnarray}
\overline{\imath \Omega_{\mathbf{q}} \mathbf{V}_{\mathbf{q}}}= \overline{\lambda_{\mathbf{q}}\mathbf{V}_{\mathbf{q}}}
\Leftrightarrow -\imath \Omega_{\mathbf{q}} \overline{\mathbf{V}_{\mathbf{q}}}=\lambda_{\mathbf{q}} \overline{\mathbf{V}_{\mathbf{q}}}
\Leftrightarrow \mathcal{K}_{\mathbf{q}}\overline{\mathbf{V}_{\mathbf{q}}}=-\lambda_{\mathbf{q}} \overline{\mathbf{V}_{\mathbf{q}}}.
\end{eqnarray}
It means that if the eigenvector $\mathbf{V}_{\mathbf{q}}$ is real, then $\lambda_{\mathbf{q}}=0$. But if $\mathbf{V}_{\mathbf{q}}$ is complex, then $-\lambda_{\mathbf{q}}$ is also an eigenvalue of the interaction matrix $\mathcal{K}_{\mathbf{q}}$. Hence, breathing lattices possess particle-hole symmetry at $\alpha=-1$ [Figs.~\ref{fig:pyrom} and \ref{fig:kagm}].\\ Furthermore $\mathcal{K}_{\mathbf{q}}$ becomes nil at the $\Gamma$ point. It means that all the bands connect at $\mathbf{q}=0$ and that the flat bands are at zero energy.\\

An interesting outcome is that the dispersive bands form Dirac cones at the $\Gamma$ point [Figs.~\ref{fig:pyrom} and \ref{fig:kagm}]. In linear order in $|\mathbf{q}|$, the dispersion relation is $\lambda^{2}_{\bf q} = -\lambda^{1}_{\bf q} = |\mathbf{q}|$ for pyrochlores and $\lambda^{2}_{\bf q} = -\lambda^{1}_{\bf q} = \sqrt{3/2}|\mathbf{q}|$ for kagomes. Please note that the presence of a single Dirac cone per Brillouin zone is allowed despite the Nielsen-Ninomiya theorem because of their co-existence with flat bands~\cite{Dagotto1986}.

\section{Ground states for $\alpha<0$}

\subsection{Breathing pyrochlore}

\begin{figure}[h]
\centering
\includegraphics[width=7.5cm]{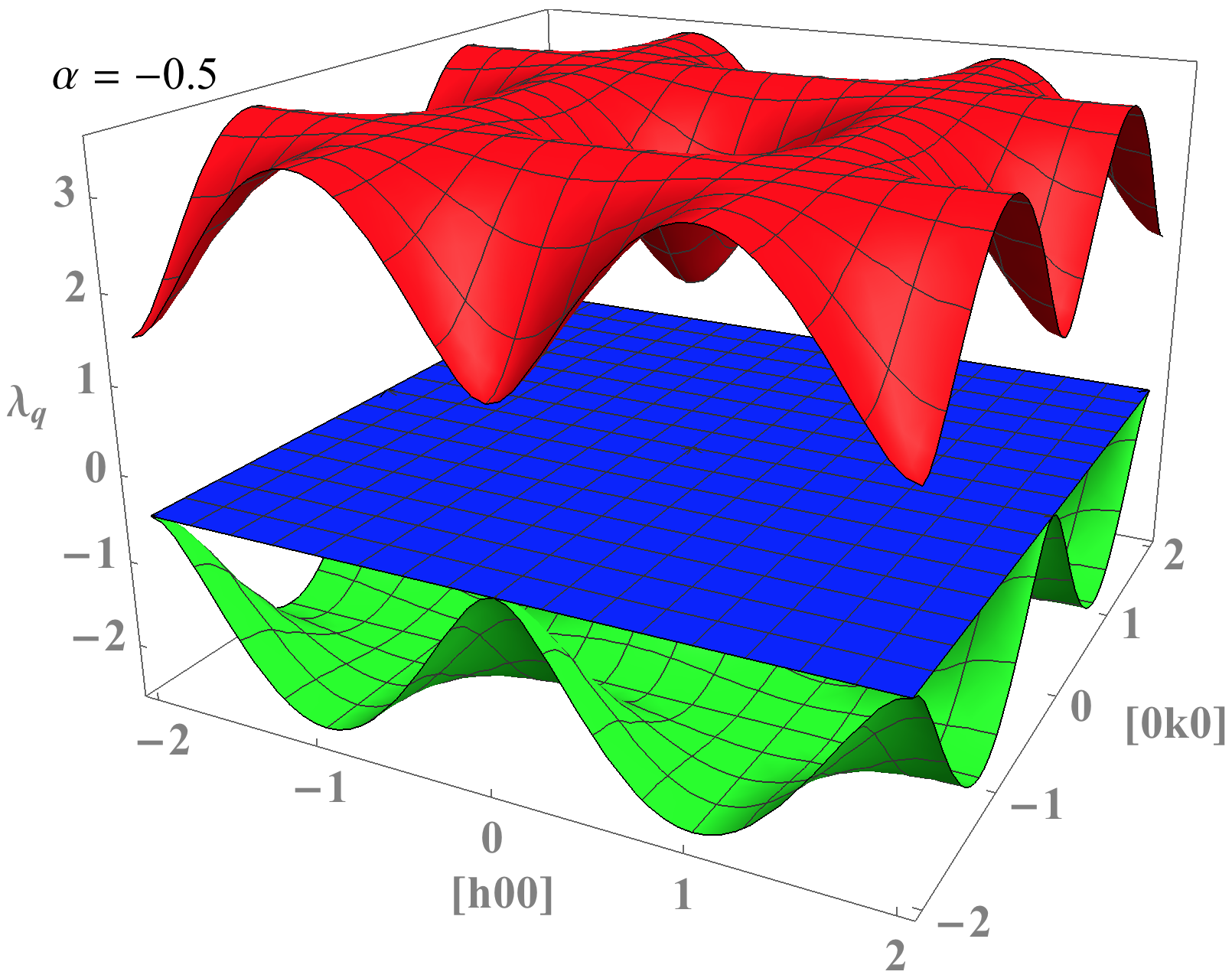}
\includegraphics[width=7.5cm]{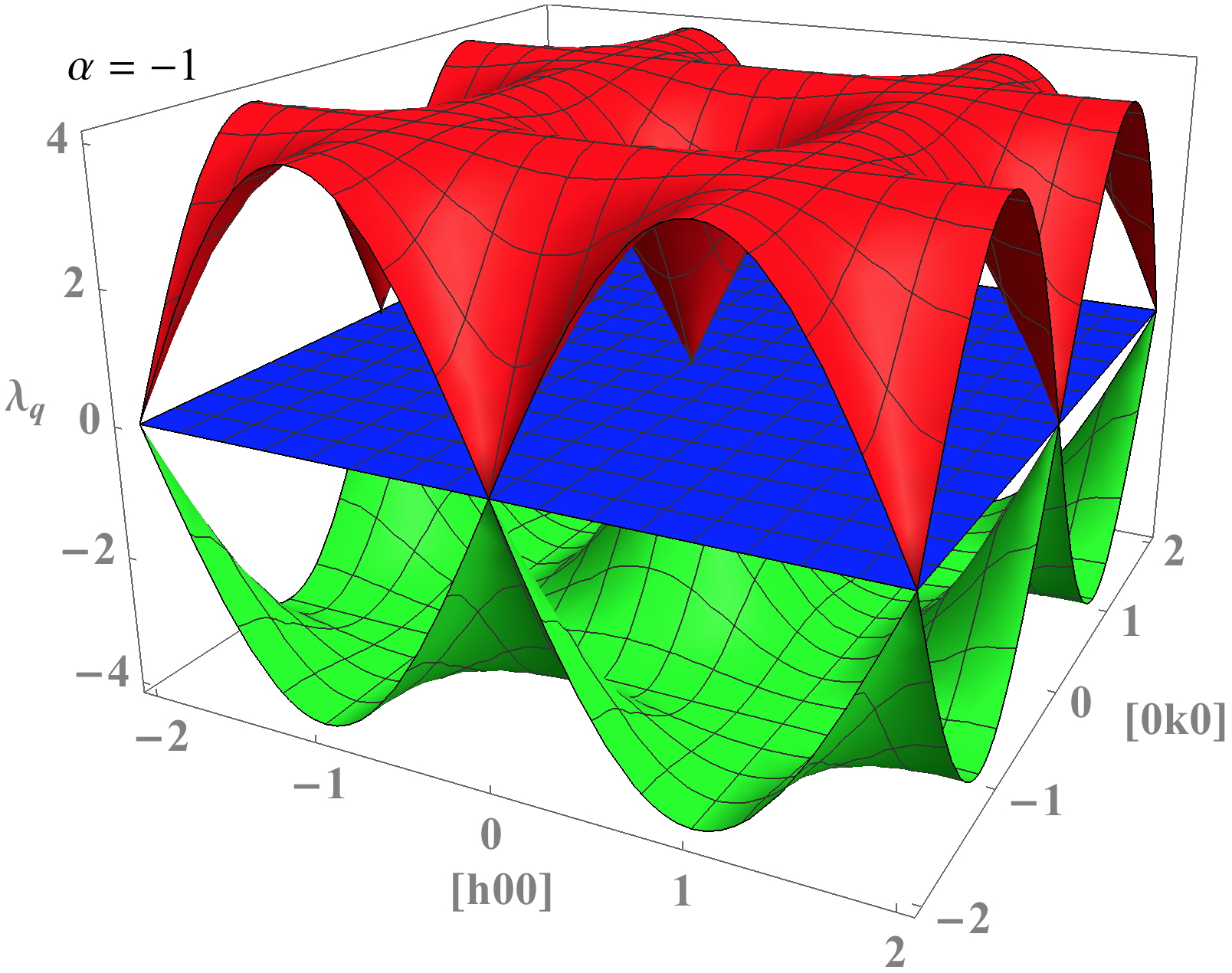}
\caption{Mode spectrum of $\mathcal{K}_{\mathbf{q}}$ for breathing pyrochlores in the [hk0] plane at $\alpha=-0.5$ and $\alpha=-1$. The [1k0] directions (and equivalent wavevectors) form lines of zero-energy modes. Wavevectors are in $2\pi$ units.}
\label{fig:pyrohk0}
\end{figure}

In agreement with the work of Benton \& Shannon~\cite{Benton2015}, we find that the classical spin liquid of the pyrochlore antiferromagnet ($\alpha=+1$) is robust for positive $\alpha$ but disappears as soon as $\alpha$ becomes negative [Figs.~\ref{fig:pyrop} and \ref{fig:pyrom}]. For $\alpha<0$ the ground-state configurations belong to one of the dispersive band, with lines of zero-energy modes along the [1k0] directions (and equivalent wavevectors) [Fig.~\ref{fig:pyrohk0}]. An order-by-disorder transition is expected to select spin configurations with [100] ordering~\cite{Benton2015}.

\subsection{Breathing kagome}

While the evolution of the band structure shares many similarities between breathing pyrochlore and kagome, their low temperature physics differ qualitatively. Indeed, there are no lines of zero-energy fluctuations in kagome for $\alpha<0$ [Fig.~\ref{fig:kagm}]. \textit{Au contraire}, the band structure shows clear minima at the corners of the Brillouin zone, which can be understood as follows.\\

\begin{figure}[b]
\centering\includegraphics[width=7cm]{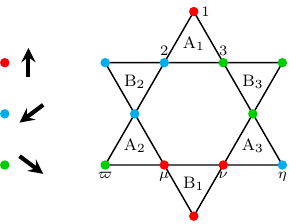}
\caption{Example of a ground-state configuration of the breathing kagome model for $\alpha<0$. Each colored dot correspond to a given spin orientation given on the left.
}
\label{fig:kagGS}
\end{figure}

In order to minimize the energy of an antiferromagnetic triangle (type A) with classical Heisenberg spins, the three spins must be coplanar with a $2\pi/3$ angle between them. For a ferromagnetic triangle (type B), spins must be collinear. Any configuration which respect these constraints over the entire lattice belongs to the ground-state manifold, irrespectively of the value of $\alpha$ as long as $\alpha<0$. Such a configuration is depicted on Fig.~\ref{fig:kagGS} and is uniquely determined once the spin configuration of one A-type triangle is given. To prove this uniqueness, let A$_{1}$ in Fig.~\ref{fig:kagGS} be this initial triangle. Without any loss of generality, its spin configuration is chosen as
\begin{eqnarray}
\mathbf{S}_{1}=\begin{pmatrix}0\\1\\0\end{pmatrix},\quad
\mathbf{S}_{2}=\dfrac{1}{2}\begin{pmatrix}-\sqrt{3}\\-1\\0\end{pmatrix},\quad
\mathbf{S}_{3}=\dfrac{1}{2}\begin{pmatrix}\sqrt{3}\\-1\\0\end{pmatrix},
\label{eq:A1}
\end{eqnarray}
whose respective color codes are red, cyan and green in Fig.~\ref{fig:kagGS}. Being ferromagnetic, the neighbouring triangles B$_{2}$ and B$_{3}$ are immediately fixed into configurations $\mathbf{S}_{2}$ and $\mathbf{S}_{3}$ respectively. On the other hand, the orientations of the four spins $\mathbf{S}_{i\in\{\varpi,\mu,\nu,\eta\}}$ are \textit{a priori} not fixed. The local antiferromagnetic constraints on triangles A$_{i\in\{2,3\}}$ indeed allow for a O(2) degree of freedom for each triangle. These continuous degrees of freedom are however constrained by the intervening ferromagnetic triangle B$_{1}$ which imposes $\mathbf{S}_{\mu}=\mathbf{S}_{\nu}$. From Eq.~(\ref{eq:A1}), this constraint can be rewritten as
\begin{eqnarray}
\mathbf{S}_{\mu}=\mathbf{S}_{\nu}
\Leftrightarrow
\begin{pmatrix}
\dfrac{\sqrt{3}}{4}(1+\sin \theta_{\mu})\\
\dfrac{1}{4}(1-3\sin \theta_{\mu})\\
\dfrac{\sqrt{3}}{2} \cos \theta_{\mu}
\end{pmatrix}
=
\begin{pmatrix}
\dfrac{\sqrt{3}}{4}(-1+\sin \theta_{\nu})\\
\dfrac{1}{4}(1+3\sin \theta_{\nu})\\
\dfrac{\sqrt{3}}{2} \cos \theta_{\nu}
\end{pmatrix}
\Leftrightarrow
\theta_{\nu}=-\theta_{\mu}=\dfrac{\pi}{2}
\Leftrightarrow
\mathbf{S}_{\mu}=\mathbf{S}_{\nu}=\mathbf{S}_{1}.
\label{eq:B1}
\end{eqnarray}
Step by step, this argument around a plaquette of six triangles can be applied to the entire lattice. The ground-state of the breathing kagome model for $\alpha<0$ is thus long-range ordered with a global degeneracy of O(3)$\times$O(2). In particular, there are no weathervane modes~\cite{Chalker1992,Zhitomirsky2008}.

If one assigns its corresponding color -- red, cyan, green -- to each of the ferromagnetic (B-type) triangles, then any ground state configuration can be seen as a three-coloring problem on the triangular lattice. The magnetization of the system is thus zero. On the other hand, the vector chirality defined as
\begin{eqnarray}
\pmb{\chi}\equiv \dfrac{1}{N} \sum_{c\,\in\{A,B\}} \left(\mathbf{S}^{1}_{c}\times\mathbf{S}^{2}_{c} + \mathbf{S}^{2}_{c}\times\mathbf{S}^{3}_{c} + \mathbf{S}^{3}_{c}\times\mathbf{S}^{1}_{c}\right)
\label{eq:vecchir}
\end{eqnarray}
is finite since all antiferromagnetic (A-type) triangles carry the same chirality while the ferromagnetic (B-type) ones bear none. Within the ground-state manifold, the vector chirality $\pmb{\chi}$ is uniformly distributed on a sphere of radius $\sqrt{3}/2$. Please note that the scalar chirality is nil since the configurations are coplanar.\\

Interestingly, if the interactions between spins are made anisotropic by favouring in-plane couplings over the out-of-plane ones (XXZ model), then the configuration of Fig.~\ref{fig:kagGS} remains one of the ground states. The only difference is that spins now lie in the kagome plane, and the ground-state manifold has $\mathbb{Z}_{2}\times$O(2) symmetry. The O(2) symmetry has the same origin as before (in-plane rotation of all spins) while the $\mathbb{Z}_{2}$ symmetry corresponds to the sign of the vector chirality $\pmb{\chi}=\pm \sqrt{3}/2\, \mathbf{e}_{z}$ where $\mathbf{e}_{z}$ is the out-of-plane unit vector.

\section{Conclusion}

We have calculated the mode spectrum of the interaction matrix $\mathcal{K}_{\mathbf{q}}$ for the breathing pyrochlore and breathing kagome lattices. While the nature and degeneracy of the flat bands are not affected by the breathing anisotropy, the shape of the dispersive bands dramatically changes as a function of $\alpha$. This is best illustrated for $\alpha=-1$ where a Dirac cone is formed at the $\Gamma$ point and the system gains particle-hole symmetry. The extensive ground-state degeneracy of the antiferromagnetic models is lifted when $\alpha<0$, in favour of lines of zero modes for breathing pyrochlore~\cite{Benton2015} and a long-range ordered ground state with finite vector chirality for breathing kagome.\\

A promising outcome of our work is the observation of a Dirac cone in a particle-hole symmetric model for $\alpha=-1$. The coexistence of Dirac nodes and flat bands has also been observed for example in a tight-binding model on the Lieb lattice with three species of spinless fermions~\cite{Palumbo15a} and on the Shastry-Sutherland lattice with Dzyaloshinskii-Moriya interactions in a field~\cite{Romanyi2015}. In both cases, it was possible to open and close the gap at the Dirac point between the flat band and \textit{both} dispersive bands by tuning either the tunneling coefficients~\cite{Palumbo15a} or the magnetic field~\cite{Romanyi2015}. It would be interesting to see if perturbations of our model could open such a gap, which would establish breathing lattices as a favourable setup for topological phenomena~\cite{Bzdusek2015}.

\ack
This work is supported by the Okinawa Institute of Science and Technology Graduate University and by KAKENHI (Nos. 26400339, 15H05852, 15K13533 and 16H04026).

\providecommand{\newblock}{}

\end{document}